\documentclass[review]{elsarticle}

\usepackage[table]{xcolor}  
\usepackage{mathtools}
\usepackage{amsmath}
\usepackage{amsmath,amssymb,amsthm}
\usepackage{graphicx}
\usepackage{subfig}
\usepackage[framemethod=tikz]{mdframed}
\usepackage{soul}

\usepackage{multirow}
\usepackage{commath}
\usepackage{url}

\usepackage{caption}
\usepackage{esvect}
\usepackage{booktabs} 
\usepackage{gensymb}
\usepackage{algcompatible}

\usepackage{algorithm}
\usepackage[noend]{algpseudocode}
\usepackage{setspace}

\makeatletter
\renewcommand{\ALG@beginalgorithmic}{\small}
\makeatother

\usepackage{tablefootnote}
\usepackage[margin=3cm]{geometry}

\makeatletter
\def\therule{\makebox[\algorithmicindent][l]{\hspace*{.5em}\vrule height .75\baselineskip depth .25\baselineskip}}%

\newtoks\therules
\therules={}
\def\appendto#1#2{\expandafter#1\expandafter{\the#1#2}}
\def\gobblefirst#1{
	#1\expandafter\expandafter\expandafter{\expandafter\@gobble\the#1}}%
\def\LState{\State\unskip\the\therules}
\def\LStatex{\Statex\unskip\the\therules}
\def\pushindent{\appendto\therules\therule}%
\def\popindent{\gobblefirst\therules}%
\def\printindent{\unskip\the\therules}%
\def\printandpush{\printindent\pushindent}%
\def\popandprint{\popindent\printindent}%

\algdef{SE}[WHILE]{While}{EndWhile}[1]
{\printandpush\algorithmicwhile\ #1\ \algorithmicdo}
{\popandprint\algorithmicend\ \algorithmicwhile}%
\algdef{SE}[FOR]{For}{EndFor}[1]
{\printandpush\algorithmicfor\ #1\ \algorithmicdo}
{\popandprint\algorithmicend\ \algorithmicfor}%
\algdef{S}[FOR]{ForAll}[1]
{\printindent\algorithmicforall\ #1\ \algorithmicdo}%
\algdef{SE}[LOOP]{Loop}{EndLoop}
{\printandpush\algorithmicloop}
{\popandprint\algorithmicend\ \algorithmicloop}%
\algdef{SE}[REPEAT]{Repeat}{Until}
{\printandpush\algorithmicrepeat}[1]
{\popandprint\algorithmicuntil\ #1}%
\algdef{SE}[IF]{If}{EndIf}[1]
{\printandpush\algorithmicif\ #1\ \algorithmicthen}
{\popandprint\algorithmicend\ \algorithmicif}%
\algdef{C}[IF]{IF}{ElsIf}[1]
{\popandprint\pushindent\algorithmicelse\ \algorithmicif\ #1\ \algorithmicthen}%
\algdef{Ce}[ELSE]{IF}{Else}{EndIf}
{\popandprint\pushindent\algorithmicelse}%
\algdef{SE}[PROCEDURE]{Procedure}{EndProcedure}[2]
{\printandpush\algorithmicprocedure\ \textproc{#1}\ifthenelse{\equal{#2}{}}{}{(#2)}}%
{\popandprint\algorithmicend\ \algorithmicprocedure}%
\algdef{SE}[FUNCTION]{Function}{EndFunction}[2]
{\printandpush\algorithmicfunction\ \textproc{#1}\ifthenelse{\equal{#2}{}}{}{(#2)}}%
{\popandprint\algorithmicend\ \algorithmicfunction}%
\makeatother

\usepackage{subfig}

\usepackage{footnote}

\makeatletter
\def\BState{\State\hskip-\ALG@thistlm}
\makeatother

\usepackage{setspace}

\usepackage{breakcites}
\usepackage{enumerate}
\usepackage{makecell}

\usepackage{float}
\usepackage{listings}

\usepackage{tikz}
\usetikzlibrary{trees}
\usepackage{paralist}
\usepackage{ragged2e}
\usepackage[framemethod=tikz]{mdframed}

\usetikzlibrary{positioning,arrows.meta}

\definecolor{arrowblue}{RGB}{98,145,224}

\usetikzlibrary{arrows.meta}
\tikzset{%
	>={Latex[width=2mm,length=2mm]},
	base/.style = {rectangle, rounded corners, 
		minimum width=11cm, minimum height=1cm,
		text centered, font=\sffamily},
	activityStarts/.style = {base, fill=blue!30},
	startstop/.style = {base, fill=red!30},
	activityRuns/.style = {base, fill=green!30},
	activityRuns2/.style = {base, fill=magenta!30},
	activityRuns3/.style = {base, fill=yellow!30},
	activityRuns4/.style = {base, fill=violet!30},
	activityRuns5/.style = {base, fill=orange!30},
	activityRuns6/.style = {base, fill=cyan!30},
	activityRuns7/.style = {base, fill=purple!30},
	activityRuns8/.style = {base, fill=gray!30},
	activityRuns9/.style = {base, fill=orange!30},
	activityRuns10/.style = {base, fill=blue!30},
	process/.style = {base, minimum width=10 cm, fill=orange!15,
		font=\ttfamily},
}
\usepackage{verbatim}
\usepackage{smartdiagram}
\usepackage{metalogo}

\usepackage{footnote}

\makeatletter
\def\BState{\State\hskip-\ALG@thistlm}
\makeatother

\usepackage{setspace}

\usepackage{breakcites}
\usepackage{enumerate}
\usepackage{makecell}

\usepackage{float}
\usepackage{listings}

\usepackage{tikz}
\usetikzlibrary{trees}
\usepackage{paralist}
\usepackage{ragged2e}
\usepackage[framemethod=tikz]{mdframed}

\usetikzlibrary{positioning,arrows.meta}

\definecolor{arrowblue}{RGB}{98,145,224}

\usepackage{hyperref}
\usepackage{cleveref}
\usepackage{setspace}
\journal{Journal of Pervasive and Mobile Computing}

\begin{document}

\begin{frontmatter}

\title{Efficient Data Perturbation for Privacy Preserving and Accurate Data Stream Mining}

\author[mymainaddress,mysecondaryaddress]{M.A.P.~Chamikara
	\corref{mycorrespondingauthor}}
\cortext[mycorrespondingauthor]{Corresponding author}
\ead{pathumchamikara.mahawagaarachchige@rmit.edu.au}

\author[mymainaddress]{P.~Bertok}
\author[mysecondaryaddress]{D.~Liu}
\author[mysecondaryaddress]{S.~Camtepe}
\author[mymainaddress]{I.~Khalil}

\address[mymainaddress]{RMIT University, Australia}
\address[mysecondaryaddress]{CSIRO Data61, Australia}

\begin{abstract}
 The widespread use of the Internet of Things (IoT) has raised many concerns, including the protection of private information. Existing privacy preservation methods cannot provide a good balance between data utility and privacy, and also have problems with efficiency and scalability. This paper proposes an efficient data stream perturbation method (named as $P^2RoCAl$). $P^2RoCAl$ offers better data utility than similar methods: classification accuracies of $P^2RoCAl$ perturbed data streams are very close to those of the original data streams. $P^2RoCAl$ also provides higher resilience against data reconstruction attacks.

\end{abstract}


\end{frontmatter}


\section{Introduction}

The Internet of Things (IoT) is becoming widely popular as it connects typical day to day devices such as kitchen appliances, cars, washing machines, headphones, wearables, etc. to the Internet to allow life activities to be more intelligent, efficient and reliable \cite{de2016iot}. IoT devices can forward a significant amount of data in streams from various sensors via long-lasting connections. IoT has revolutionized many fields including health-care, wellbeing applications, social life, environment monitoring, transportation, and energy. The availability of low-cost pervasive sensing devices has enabled IoT to grow in an ever-increasing manner, and IoT sensor streams have become a vital source of big data \cite{de2016iot,balandina2015iot}.

The vast diversity of IoT devices introduces many challenges \cite{gazis2015short}, among which the most important ones are (i) effective data collection, (ii) efficient data processing, and (iii) privacy protection and security \cite{shang2016challenges}. The availability and increased accessibility of IoT sensors often introduce the risk of privacy breach for individuals. Pervasive data collection, such as crowd sensing applications, may involve sensitive personal information including lifestyle, habits and personal preferences, and accessibility to such information by a third-party may raise privacy concerns \cite{rehman2016security}. It is important to share IoT streams (or big data in general) so that authorized third parties can make valuable decisions, but information extracted from these streams should not be linkable to individuals.

Disclosure control of microdata refers to the process of applying different privacy-preserving mechanisms to the data before releasing them for analysis~\cite{bethlehem1990disclosure}. Privacy-preserving data mining (PPDM) has been widely investigated for big data, and similar approaches to IoT streams are now emerging. Privacy-preserving methods are struggling to achieve higher accuracy~\cite{aggarwal2008general}, and the exponential growth in IoT-sourced data streams adds another layer of difficulties~\cite{lu2014toward}. The popular  PPDM  techniques are either not scalable or not efficient enough to deal with this, and privacy preservation of data streams is still a significant challenge. Technological advancements in data storage have eased the burden of storing large chunks of data generated by a variety of sources \cite{hashem2015rise}. In addition to the issues posed by big data, IoT faces the problem of efficient data processing, as the devices are most often resource constrained. Therefore, the efficiency of privacy preservation techniques is essential. The unpredictability and diversity of data streams form a challenging environment for privacy-preserving algorithms that want to achieve high accuracy. 

Data perturbation is a privacy preservation technique that alters the values of data elements in a database to maintain individual record confidentiality~\cite{kargupta2003privacy}. Among PPDM techniques, data perturbation is relatively simple. Additive perturbation \cite{muralidhar1999general},  \cite{reiss1980practical}, random rotation \cite{chen2005random}, geometric perturbation \cite{chen2011geometric}, microaggregation \cite{domingo2002practical} and condensation \cite{aggarwal2004condensation} are some of the existing perturbation methods. Privacy preservation of IoT data streams presents several additional challenges, as the data is released incrementally, endlessly, and its fast nature hinders the possibility of using historical information. Differential privacy (DP) has attracted attention due to the level of privacy guarantee it provides~\cite{dwork2006calibrating,dwork2009differential, mohammed2011differentially,friedman2010data,soria2017individual}. DP is a privacy model, similar to $k-anonymity$~\cite{niu2014achieving},  $l-diversity$~\cite{machanavajjhala2006diversity}, which defines a strong privacy guarantee over data. But deploying methods to achieve DP over data streams is challenging due to the limit of privacy budget offered by it ~\cite{yang2012differential}.  Prominent attempts of privacy preservation of data streams include anonymization~\cite{cao2011castle}, randomization, microaggregation and data condensation \cite{aggarwal2008general,domingo2017steered}. However, these methods have not been able to balance privacy and accuracy effectively: when one is preserved, the other seems to suffer.

The main contribution of this paper is an efficient and secure algorithm that can be used for perturbing high-speed data streams such as those produced by IoT devices. Fast execution with predictable execution time guarantees the capability of working with continuously growing big data. The method first conducts homogeneous group formation (based on tuple distances) and uses group properties to generate a rotation matrix to perform rotation perturbation upon each group. The efficiency of the proposed method was proven by testing it on generic datasets retrieved from the UCI data repository \footnote{https://archive.ics.uci.edu/ml/index.php}. Classification accuracy of the perturbed datasets was also proven by using different classification algorithms. The results indicate that the proposed method is very effective in privacy-preserving data stream classification. The proposed algorithm's privacy protection was demonstrated against several, often-used attack methods, including naive estimation \cite{aggarwal2008general}, known I/O \cite{aggarwal2008general} attacks and Independent Component Analysis (ICA)  \cite{aggarwal2008general}. A comparison with random rotation perturbation \cite{chen2005random} and data condensation \cite{chen2011geometric} indicated the proposed method's superiority in both classification accuracy and data privacy. Clear advantages of the proposed scheme in efficiency, accuracy and data privacy make it an excellent solution for data streams demanding on-line processing in resource limited environments. 

The remainder of this paper is organized in the following manner. Section \ref{literaturereview} provides a literature review on the existing related methods. Section \ref{method} provides the methodology employed by $P^2RoCAl$ along with its variations. Section \ref{results} presents the results produced by $P^2RoCAl$ in the sense of classification accuracy, attack resilience, time complexity, and scalability. A comprehensive discussion on the results is provided under Section \ref{discussion} with some future directions. The paper is concluded in Section \ref{conclusion}.

\section{Literature Review}
\label{literaturereview}
IoT enables the interaction of heterogeneous information systems to deliver important, diversified services to users. IoT devices are used in many domains such as health-care, smart cities, and wearables \cite{de2016iot}. The explosive growth of IoT has produced a vulnerable medium that may leak sensitive data to unwanted third parties. IoT often includes wireless sensor networks to collect and process data, hence generates incremental data streams that eventually result in big data. IoT often brings up many security and privacy issues such as confidentiality, authentication and authorization \cite{balandina2015iot}. But, it is also important that authorized third parties be able to retrieve IoT stream data and big data, e.g. to generate valuable insights using data mining techniques. Different attempts in the literature tried to impose privacy on IoT data, such as controlling access over authentication \cite{bertino2016data}, Attribute-Based Encryption \cite{wang2014performance}, temporal and location-based access control \cite{bertino2016data} and employing constraint-based protocols \cite{kirkham2015privacy} are a few examples of such methods. Yabo Xu et al. have adapted Naive Bayesian Classifier for private data streams, but their method is not suitable for generic data stream classification \cite{xu2008privacy}. Feifei Li et al. derived a method to efficiently and effectively track the correlation and autocorrelation structure of multivariate streams and leverage it to add noise to preserve privacy. However, the method is vulnerable to principal component analysis-based attacks \cite{li2007hiding}. Josep Domingo-Ferrer et al. proposed a  method named Steered Microaggregation that can be used to anonymize a data stream to achieve k-anonymity. But the problem of information leak inherent to k-anonymity over high dimensional data can be a shortcoming of the method~\cite{domingo2017steered}. Aggarwal C.C. et al. proposed a condensation based privacy-preserving method for data mining. The method is scalable and efficient, but its ability to maintain a good balance between privacy and utility is questionable. When the method parameters are set to achieve high accuracy (using small spatial locality), the privacy of the data often suffers \cite{ aggarwal2008static}. Condensation can be considered as a descendant of microaggregation which follows a similar mechanism of homogeneous group formation. Both microaggregation and condensation use the concept of within-group homogeneity. The main difference between them is that, while in microaggregation a single central tendency measure is used, condensation uses a covariance matrix based mechanism~\cite{aggarwal2004condensation,domingo2002practical}.

Privacy can have different meanings and definitions. From our perspective, privacy can be considered as ``Controlled Information Release'' \cite{bertino2008survey}. The advancements in data mining and dissemination methods increased the demand for preservation of privacy  \cite{verykios2004state}. As a result, privacy preservation became an important prerequisite of any data mining system, and literature shows many attempts to address the related issues \cite{aldeen2015comprehensive}. Privacy-preserving techniques can be classified as data distribution scenarios (centralized or distributed), data modification methods (perturbation, encryption, generalization, etc.), data mining algorithms, and data or rule hiding methods \cite{verykios2004state}. Data perturbation comes under data modification methods, where the data are subjected to modification using different approaches such as noise addition \cite{muralidhar1999general}, microaggregation \cite{domingo2002practical} and randomization. Perturbation attracted attention due to its relative simplicity and efficiency compared to other PPDM methods \cite{aldeen2015comprehensive}. Data perturbation methods can be classified into input perturbation and output perturbation. The latter is based on the concepts of noise addition and rule hiding while the former is performed either by the addition of noise \cite{muralidhar1999general} or multiplication by noise \cite{liu2006random}. Input perturbation can  be divided further into unidimensional perturbation and multidimensional perturbation  \cite{agrawal2000privacy,datta2004random,liu2006random,zhong2012mu}. Additive perturbation \cite{muralidhar1999general}, randomized response \cite{du2003using}, and swapping \cite{estivill1999data} are types of unidimensional input perturbation;  microaggregation possesses both unidimensional and multidimensional perturbation capabilities \cite{domingo2002practical}, whereas condensation \cite{aggarwal2004condensation}, random rotation \cite{chen2005random}, geometric perturbation \cite{chen2011geometric}, random projection \cite{liu2006random}, and hybrid perturbation are types of multidimensional perturbation ~\cite{aldeen2015comprehensive}.

A privacy model should  identify the limits of private information protection/disclosure  ~\cite{machanavajjhala2015designing}. Earlier privacy models, such as $k-anonymity$~\cite{niu2014achieving}, $l-diversity$~\cite{machanavajjhala2006diversity}, $(\alpha, k)-anonymity$~\cite{wong2006alpha}, $t-closeness$~\cite{li2007t} show vulnerability to different attacks, e.g. minimality~\cite{zhang2007information}, composition~\cite{ganta2008composition} and foreground knowledge~\cite{wong2011can} attacks. Later, it was proven that many of these models face challenges in regards to the curse of dimensionality~\cite{aggarwal2005k,cao2011castle}. For example in the cases of big data and data streams, the methods that try to satisfy these privacy models tend to have higher privacy leak~\cite{aggarwal2005k}. Differential privacy (DP) is another privacy model that provides a strong privacy guarantee. Global differential privacy (GDP) and local differential privacy (LDP) are the two models that can be used to achieve DP over data.   GDP is based on output perturbation where the method and amount of data perturbation are determined based on the query outputs.  GDP is also called the trusted curator model, where the analyst would only be able to run queries on the database whereas in local differential privacy (LDP) input perturbation (e.g. using randomized response~\cite{fox2015randomized}) is applied to facilitate full/ partial data release. In LDP the analysts can run their analysis directly upon the perturbed data  ~\cite{dwork2008differential,kairouz2014extremal}, e.g. increasing information content via injection of additional noise to minimize privacy leakage~\cite{tang2017privacy}. However, GDP and LDP show poor performance for small datasets, as accurate estimation of the statistics shows poor results over heavily noisy data. Furthermore, existing LDP algorithms involve a significant amount of noise addition/ randomization on input data, resulting in a lower level of utility for data streams. As the data stream's lifespan extends, the GDP and LDP processes will have to be restarted when the privacy budget is reached. This drastically reduces the utility of DP in the perspective of data streams~\cite{yang2012differential}.  Utility and privacy often appear as conflicting factors, and improved privacy is frequently accompanied by reduced utility~\cite{mivule2013comparative}.

The most relevant to the approach we propose are two perturbation methods: data condensation and rotation perturbation. In data condensation, the data are divided into multiple homogeneous groups of predefined size in such a way that the difference between the records in a particular group is minimal,  and a certain level of statistical information about different records is maintained. Then the sanitized data are generated using the uniform random distribution based on the eigenvectors which are generated using the eigendecomposition  of the characteristic covariance matrices of each homogeneous group \cite{aggarwal2004condensation}. Although condensation is a good contender among privacy preservation algorithms, in some instances the condensation approach fails to provide enough privacy, especially when the spatial locality of subgroups are small, which can be a characteristic feature of data streams. On the other hand, when spatial locality is large, the quality of the dataset drastically reduces \cite{chen2011geometric}, and hence the accuracy decreases. Random rotation perturbation is a matrix multiplicative \cite{liu2007multiplicative} data perturbation method, in which the original data matrix is multiplied using a random rotation matrix that has the properties of an orthogonal matrix. The application of rotation is iterated until the algorithm converges at the desired privacy level \cite{chen2005random}. Due to the isometric nature of transformations, random rotation perturbation is capable of preserving the distances between the tuples in the original datasets \cite{chen2005random}, and therefore it provides high utility towards classification and clustering. However, due to the recursive nature of random data generation, this method consumes an extensive amount of computer resources, making it difficult to work with extensively large datasets.

Literature shows many attack methods that can be used to re-identify individuals in a sanitized database. It has been noted that different methods are vulnerable to different types of attacks. There are many types of attack methods such as principal component analysis \cite{wold1987principal}, maximum likelihood estimation \cite{scholz1985maximum}, known I/O attack \cite{aggarwal2008general}, ICA attack \cite{chen2005privacy} and known sample attack \cite{aggarwal2008general}. These attacks basically try to regenerate the original dataset. For example, additive perturbation can be attacked using different data/noise reconstruction algorithms that are based on concepts such as principal component analysis \cite{huang2005deriving}, maximum likelihood estimation \cite{huang2005deriving}. Multiplicative data perturbation methods can be exploited using known input/output (I/O) attacks, known sample attacks, and ICA attacks.

Many of the previously proposed privacy preservation methods, including data perturbation, perform poorly when high dimensional datasets are introduced. Even when the performance is good for low dimensional data, as the number of attributes and number of instances increase, the amount of necessary computer resources increase exponentially \cite{machanavajjhala2006diversity, chen2005random,chen2011geometric}. This is often called ``the curse of dimensionality'', and not only makes the perturbation process extensive, it also provides extra information to attackers. The higher the dimensions  in the datasets, the easier attackers can use background knowledge to determine the identity of individuals \cite{aggarwal2005k}.

Literature shows a paucity of efficient privacy preservation methods scalable enough to handle the exponentially growing databases and data streams such as IoT stream data. It can also be noted that the existing methods have problems with utility, level of uncertainty, and low level of resilience. A new data stream perturbation method that is scalable, efficient, and robust would overcome the existing issues of the past PPDM methods and provide a solution towards large-scale privacy-preserving data stream and big data mining.

\section{Method}
\label{method}
\begin{figure}[H]
	\centering
	\scalebox{0.8}{
		\includegraphics[width=1\textwidth, trim=0cm 0cm 0cm
		0cm]{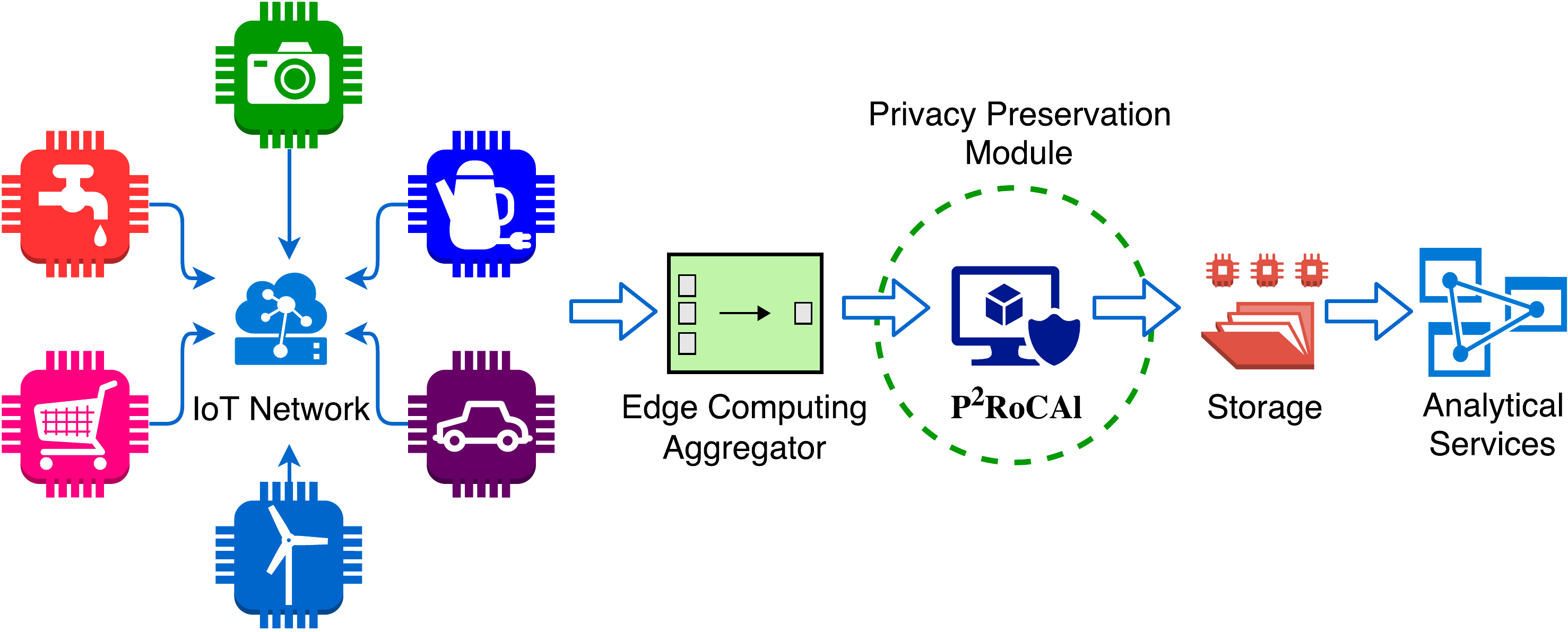}
	}
	\caption{Application of $P^2RoCAl$ in the general purpose data flow of IoT data streams.}
	\label{iotposition}
\end{figure}

This section provides a comprehensive description of the proposed method named as $P^2RoCAl$: Privacy-Preserving Rotation based Condensation Algorithm. $P^2RoCAl$  is a privacy preservation algorithm for data streams and big data. It uses the properties of data condensation and rotation perturbation and combines their qualities, the efficiency of condensation and accuracy of rotation perturbation.  $P^2RoCAl$ is designed to perturb data before the storage phase in the general purpose data flow of IoT data streams as depicted in Figure \ref{iotposition}. That is, the stored data has already gone through the privacy preservation process of $P^2RoCAl$.

\begin{figure}[H]
	\centering
	\begin{tikzpicture}[node distance=1.8cm,align=center]
	\node (start)             [activityStarts,scale=1]{Accept the dataset (with $m$ tuples and $n$ attributes) to be sanitized};
	\node (activityRuns1)      [activityRuns, below of=start,scale=1]{Accept the input value for k or k$'$};
	\node (activityRuns2)      [activityRuns2, below of=activityRuns1,scale=1]{Conduct the clustering process \\ according to Algorithm \ref{datclust}};
	\node (activityRuns3)      [activityRuns3, below of=activityRuns2,scale=1]{Generate covariance matrices of each group $C(G_i)$};
	\node (activityRuns4)      [activityRuns4, below of=activityRuns3,scale=1]{Determine the eigenvectors of each covariance matrix by decomposing \\ $C(G_i)$  in the following form:
		$C(G_i)=P(G_i)\times \Delta(G_i)\times P(G_i)^T$ };
	\node (activityRuns5)      [activityRuns5, below of=activityRuns4,scale=1]{Generate $RP(G_i)$ using random column shuffle};
	\node (activityRuns6)      [activityRuns6, below of=activityRuns5,scale=1]{Multiply the records in each group\\ using the corresponding $RP(G_i)$ of that group};
	\node (activityRuns7)      [activityRuns7, below of=activityRuns6,scale=1]{Merge the rotated groups};
	\node (activityRuns8)      [activityRuns8, below of=activityRuns7,scale=1]{Shuffle the tuples of the dataset};
	{onStop()};
	\node (ActivityDestroyed) [startstop, below of=activityRuns8,scale=1]{Release the final dataset}; 
	\draw[->]             (start) -- (activityRuns1);
	\draw[->]     (activityRuns1) -- (activityRuns2);
	\draw[->]     (activityRuns2) -- (activityRuns3);
	\draw[->]     (activityRuns3) -- (activityRuns4);
	\draw[->]     (activityRuns4) -- (activityRuns5);
	\draw[->]     (activityRuns5) -- (activityRuns6);
	\draw[->]     (activityRuns6) -- (activityRuns7);
	\draw[->]     (activityRuns7) -- (activityRuns8);
	\draw[->]     (activityRuns8) -- (ActivityDestroyed);
	\end{tikzpicture}
	\caption{Perturbation algorithm for a static database}
	\label{algoflow}
\end{figure}
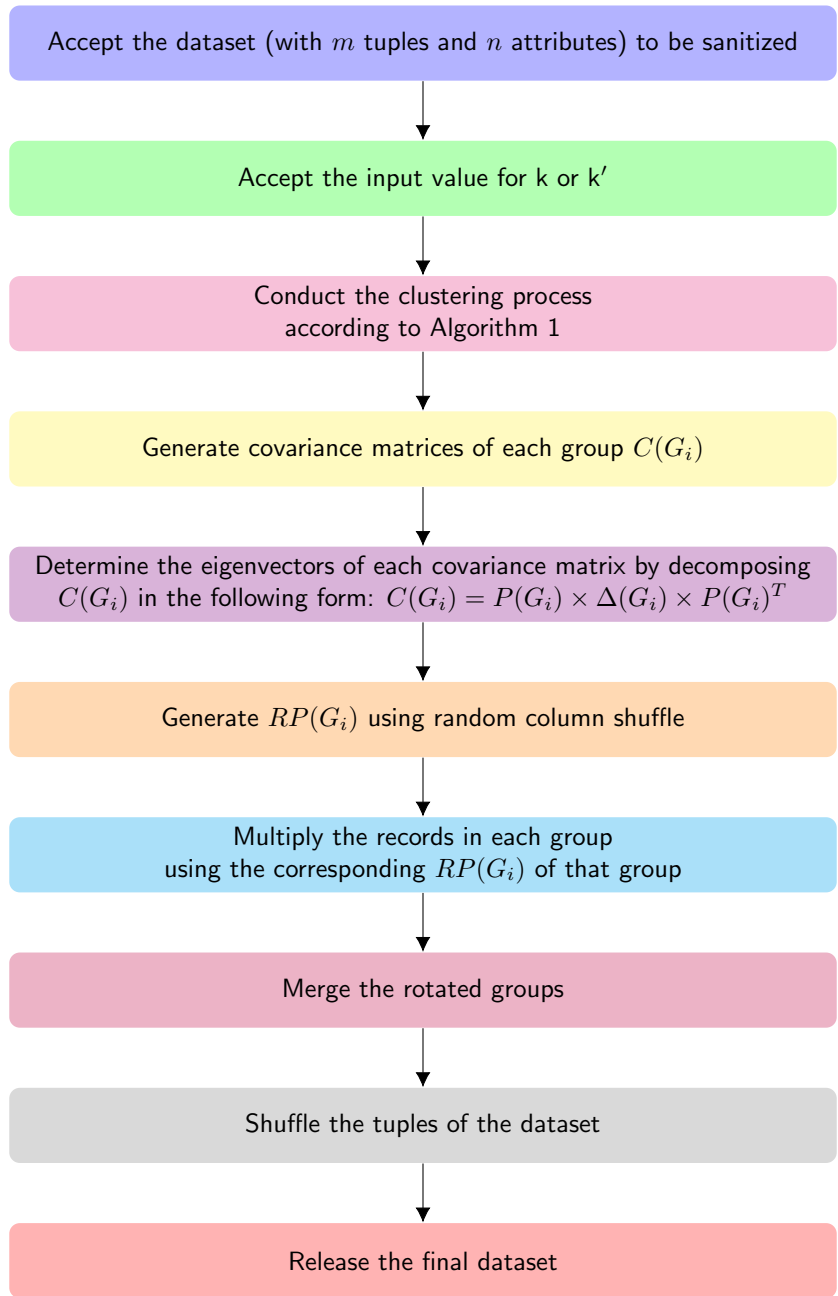

$P^2RoCAl$ initially clusters the data into multiple homogeneous groups (grouping and clustering will be used interchangeably throughout the paper with the same meaning). Data processing for privacy preservation is then imposed on the fixed-size data chunks that are dynamically maintained. Next, the covariance matrix of each group is generated using group statistics. After that, the covariance matrices are used to generate the corresponding geometric rotational groups. The rotated groups are then merged, and the tuples are randomly shuffled and released. The main steps of the proposed algorithm are presented in Figure \ref{algoflow} and Algorithm \ref{staticdata}. The initial data grouping process can be conducted using Algorithm \ref{datclust}. Parameter $k$ (number of groups/clusters) or $k'$ (number of tuples in a group/cluster) in Algorithm \ref{datclust} is accepted as a user input prior to the data grouping process. After accepting the input dataset $D$ and the initial $k$ or $k'$, the clustering/grouping of the dataset $D$ is conducted using Algorithm \ref{datclust}. Here each group/cluster is considered to be a condensed group containing the records $\{X_1, X_2,..., X_k'\}$. Next, we generate the covariance matrix $C(G_i)$ that corresponds to each group/cluster. In order to do so, we need to maintain the following information as characterized by data condensation \cite{aggarwal2004condensation}.

\begin{itemize}
	\item For each attribute, the sum of corresponding values.
	\item For each pair of attributes, the sum of the product of corresponding attribute values.
\end{itemize}

\begin{algorithm}[H]
	\caption{Data clustering/grouping process}\label{datclust}
	\begin{spacing}{1.2}
		\begin{algorithmic}[1]
			\linespread{1}
			
			\Statex\textbf{Inputs} :
			
			\begin{tabular}{l c l} 
				$D$  & $\ \gets $ & original dataset with $m$ tuples and $n$ attributes\\
				$k$  & $\ \gets $ & number of groups/clusters ($flag=1$)\\
				& or&\\
				$k'$  & $\ \gets $ & number of tuples/instances in one cluster/group  ($flag=0$)\\
			\end{tabular}
			
			\Statex \textbf{Outputs}:
			
			\begin{tabular}{ l c l } 
				$G$  & $\ \gets $ & homogeneous groups  \\
			\end{tabular}
			
			\If {$k > n$ or $k'< 2$}
			\LState $stop$
			\EndIf
			\If {$flag==1$}
			\LState conduct $k$-means clustering on the dataset to form the $k$ number of groups ($G_1, G_2,..,G_k$)
			\Else \textbf{ if} {$flag==0$} \textbf{then}
            \LState i=1
			\Repeat 
			\LState randomly sample a tuple ($X_i$) from $D$
			\LState select the closest $k'-1$ and $X_i$ from $D$ to form a cluster, $G_i$
			\LState remove the selected $k'$ tuples from $D$
            \LState i=i+1
			\Until {$D$ is empty}
			\EndIf			
			\Statex \textbf{End Algorithm}
		\end{algorithmic}
	\end{spacing}
\end{algorithm}

After generating each covariance matrix, the eigenvectors of each covariance matrix are determined by decomposing $C(G_i)$ according to Equation \ref{eiequ}. Here, the columns of $P(G_i)$ represent the eigenvectors of covariance matrix $ C(G_i)$. Since the matrix is positive semi-definite, the corresponding eigenvectors form an orthonormal axis system. Hence, the resulting matrix of eigenvectors ($P(G_i)$) of a particular covariance matrix, which corresponds to a homogeneous group, has the properties of an orthogonal matrix where columns and rows are orthonormal. Therefore, $P(G_i)$ preserves the relationship $P(G_i)\times P(G_i)^T=P(G_i)^T\times P(G_i)=I$ where $P(G_i)^T$ is the transpose matrix of $P(G_i)$ and $I$ is the identity matrix. This implies that $P(G_i)$ of a particular homogeneous group has all the properties of a rotation matrix. It was also proven that the resulting matrix is still an orthonormal matrix, although the order of the rows or the columns of the orthonormal matrix is changed \cite{chen2011geometric}. Hence the column-permuted matrix of the resulting rotation matrix will also behave as a rotation matrix. This property was next used to randomize the process of rotation perturbation by generating randomized matrix $RP(G_i)$ by a random column shuffle of $P(G_i)$.

\begin{equation}\label{eiequ}
C(G_i)=P(G_i)\times \Delta(G_i)\times P(G_i)^T
\end{equation}

\subsection{Algorithm for static datasets}

Algorithm \ref{staticdata} and the Figure \ref{algoflow} show how the perturbation is conducted on a static dataset where the whole dataset is fed to the algorithm once. At the end of the group rotations, the rotated groups will be merged, and the tuples will be randomly swapped (shuffled) to increase randomness in the final dataset to improve data privacy. Flag values of Algorithm \ref{staticdata} represent the two configuration selections for the two types of clustering/grouping possible. If the user provides $k$ (number of groups/clusters) as the input value, $flag$ will be set to 1, and the grouping will be conducted using $k-means$ clustering as fixed in Algorithm \ref{datclust}. If $k'$  (number of tuples in a group/cluster) is selected as the input, $flag$ will be set to 0 (which is the default setting). The grouping is done using the random clustering process as fixed in Algorithm \ref{datclust} under the setting of $flag=0$. In case of group size of only one tuple, the rotation will be applied according to the configurations of the closest previous group with more than one tuple. This step is used to increase the effect of perturbation, as for a group of one tuple, the result can be a rotation matrix that is equal to the identity matrix $I$. The multiplication of a vector and the identity matrix results in the same vector generating zero perturbation effect on the initial vector. This can be avoided using the perturbation process correction done in line \ref{stepref} of Algorithm \ref{staticdata} as explained above.

\begin{algorithm}[H]
	\caption{$P^2RoCAl$ for Static Datasets}\label{staticdata}
	\begin{spacing}{1.2}
		\begin{algorithmic}[1]
			\linespread{1}
			
			\Statex\textbf{Inputs} :
			
			\begin{tabular}{l c l} 
				$D$  & $\ \gets $ & original dataset with $m$ tuples and $n$ attributes\\
				$k$  & $\ \gets $ & number of groups/clusters ($flag=1$)\\
				& or&\\
				$k'$  & $\ \gets $ & number of tuples/instances in one cluster/group  ($flag=0$)\\
			\end{tabular}
			
			\Statex \textbf{Outputs}:
			
			\begin{tabular}{ l c l } 
				$D^{p}$  & $\ \gets $ & perturbed dataset  \\
			\end{tabular}
		
			\LState conduct clustering/grouping on the dataset using Algorithm \ref{datclust} \label{line1}
			\For{\textbf{each} group/cluster $ G_i $ }
			\LState generate $C(G_i) $ 
			\If {$i^{th} group size > 1 $}
			\LState $C(G_i)=P(G_i)\times \Delta(G_i)\times P(G_i)^T$ \label{line5}
			\LState generate $RP(G_i)$ using random column shuffle
			\LState $D^p(G_i)=(RP(G_i)\times D(G_i)^T)^T$ \label{line7}
			\Else
			\LState choose the last rotation matrix, $RP(G_l)$ of the group  \label{stepref}
			\LStatex with number of tuples greater than 2 
			\LState $D^p(G_i)=(RP(G_l)\times D(G_i)^T)^T$ \label{line10}
			\EndIf		
			\EndFor
			\LState $D^p=merge(D^p(G_1),D^p(G_2),\dots, D^p(G_n)) $ 
			\LState randomly swap the tuples of $D^p$ 			
			\State release $\ D^{p} $			
			\Statex \textbf{End Algorithm}
		\end{algorithmic}
	\end{spacing}
\end{algorithm}

\subsection{Algorithm for data streams}

\begin{figure}[H]
	\centering	
 	\scalebox{0.9}{
	\begin{tikzpicture}[node distance=1.8cm,align=center]
	\node (start)             [activityStarts,scale=1]{Accept the input values $k$ or $k'$\\ and $t$ threshold for data release \\($t$ is the number of data chunks of size $l$ to be released)};
	\node (activityRuns1)      [activityRuns, below of=start,scale=1]{Accept $l$ number of tuples from the data stream \\ ($l$ the buffer size where, $l>2\times k$)};
	\node (activityRuns2)      [activityRuns2, below of=activityRuns1,scale=1]{\ Conduct the clustering process \\according to Algorithm \ref{datclust}};
	\node (activityRuns3)      [activityRuns3, below of=activityRuns2,scale=1]{Generate covariance matrices of each group $C(G_i)$};
	\node (activityRuns4)      [activityRuns4, below of=activityRuns3,scale=1]{Determine the eigenvectors of each covariance matrix by decomposing \\ $C(G_i)$  in the following form:
		$C(G_i)=P(G_i)\times \Delta(G_i)\times P(G_i)^T$ };
	\node (activityRuns6)      [activityRuns6, below of=activityRuns4,scale=1]{Generate $RP(G_i)$ using random column shuffle};
	\node (activityRuns7)      [activityRuns7, below of=activityRuns6,scale=1]{Multiply the records in each group\\ using the corresponding reordered $RP(G_i)$ of that group};
	\node (activityRuns8)      [activityRuns8, below of=activityRuns7,scale=1]{Merge the rotated groups};
	\node (activityRuns9)      [activityRuns9, below of=activityRuns8,scale=1]{Shuffle the tuples of the dataset (of size $l$)};
	\node (activityRuns10)      [activityRuns10, below of=activityRuns9,scale=1]{Buffer the perturbed group (of size $l$)};
	{onStop()};
	\node (ActivityDestroyed) [startstop, below of=activityRuns10,scale=1]{Release the merged dataset of size $t\times l$ when the threshold $t$ is reached};  
	\draw[->]             (start) -- (activityRuns1);
	\draw[->]     (activityRuns1) -- (activityRuns2);
	\draw[->]     (activityRuns2) -- (activityRuns3);
	\draw[->]     (activityRuns3) -- (activityRuns4);
	\draw[->]     (activityRuns4) -- (activityRuns6);
	\draw[->]     (activityRuns6) -- (activityRuns7);
	\draw[->]     (activityRuns7) -- (activityRuns8);
	\draw[->]     (activityRuns8) -- (activityRuns9);
	\draw[->]     (activityRuns9) -- (activityRuns10);
	\draw[->]     (activityRuns10) -- (ActivityDestroyed);
	\draw[->] (activityRuns10) -- ++(6.5,0) -- ++(0,12.4) -- ++(0,2) --                
	      node[xshift=2cm,yshift=-3cm, text width=3cm]
	      {Repeat until the data stream is stopped/paused}(activityRuns1);
	\end{tikzpicture}
	}
	\caption{Perturbation algorithm for a data stream}
	\label{dynamicalgoflow}
\end{figure}
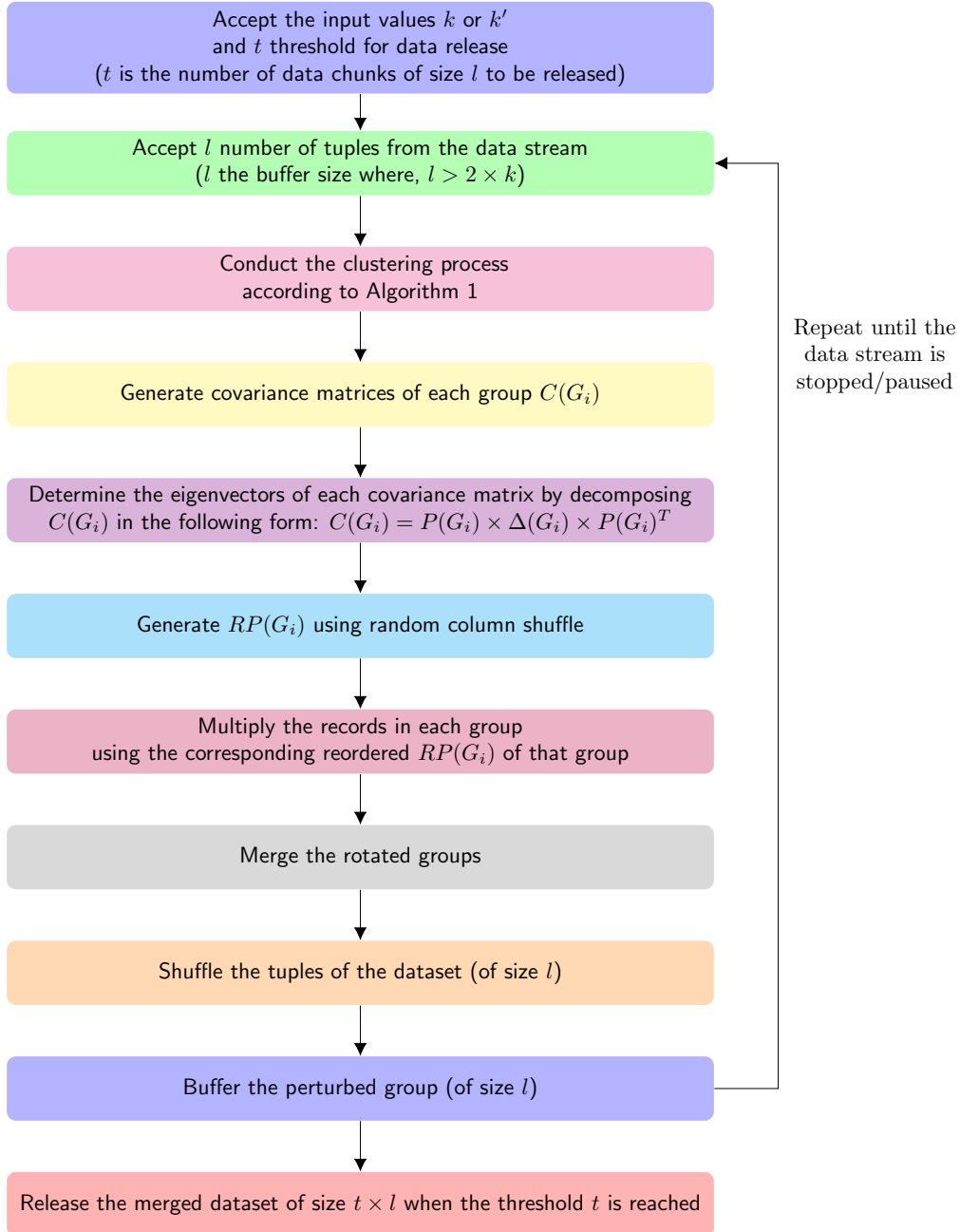

In the case of data streams, the algorithm accepts a buffer size of $l$ and a threshold ($t$) for data release where $t$ is the number of data chunks of size $l$ to be released before the data stream is paused or stopped as denoted in Algorithm \ref{datastreams} and Figure \ref{dynamicalgoflow}. The algorithm assumes that the minimum group size is two tuples, that is, the constraint of $l<2\times k$ is imposed. This constraint is to avoid the possibility of ending up with $k$ number of groups where some groups will have one tuple, and some groups will have none. 

Data streams often grow ad infinitum incrementally. This makes the process of privacy preservation complex. To reduce the complexity in the dynamic setting, a data buffer of size $l$ is dynamically maintained, as shown in Algorithm \ref{datastreams}.  This data buffer of size $l$ enables generating clusters with better homogeneity (due to the static nature of clustering).  As the algorithm executes, it will accept and buffer $l$ number of tuples and conduct the perturbation of the buffered data. After perturbation, the buffered perturbed data is released and merged into the current data chunk, and the buffer can accept the next set of $l$ tuples. Having a threshold of $t$ avoids the possibility of waiting for a data release for an unlimited amount of time.

\begin{algorithm}[H]
	\caption{$P^2RoCAl$ for Data Streams}\label{datastreams}
	\begin{spacing}{1.2}
		\begin{algorithmic}[1]
			\linespread{1}
			
			\Statex\textbf{Inputs} :
			
			\begin{tabular}{l c l} 
				$D$  & $\ \gets $ & original dataset stream with $n$ attributes\\
				$l$  & $\ \gets $ & data buffer size\\
                $t$  & $\ \gets $ & threshold for data release (number of data chunks of size $l$ to be released\\ 
                & & before the data stream stops)\\
				$k$  & $\ \gets $ & number of groups/clusters ($flag=1$)\\
				& or&\\
				$k'$  & $\ \gets $ & number of tuples/instances in one cluster/group  ($flag=0$)\\
			\end{tabular}
			
			\Statex \textbf{Outputs}:
			
			\begin{tabular}{ l c l } 
				$D^{p}$  & $\ \gets $ & perturbed dataset  \\
			\end{tabular}
			\State rep=0
			\If {$l<2\times k$}
			\LState $stop$
			\EndIf
			
			\Repeat
            \LState receive $l$ number of tuples from the data stream
            \LState rep=rep+1
			\LState conduct clustering/grouping on the dataset using Algorithm \ref{datclust}
			\For{\textbf{each} group/cluster $ G_i $ }
			\LState generate $C(G_i) $ 
			\If {$i^{th} group size > 1 $}
			\LState $C(G_i)=P(G_i)\times \Delta(G_i)\times P(G_i)^T$
			\LState generate $RP(G_i)$ using random column shuffle
			\LState $D^p(G_i)=(RP(G_i)\times D(G_i)^T)^T$ 
			\Else
			\LState choose the last rotation matrix, $RP(G_l)$ of the group 
			\LStatex with number of tuples greater than 2
			\LState $D^p(G_i)=(RP(G_l)\times D(G_i)^T)^T$
			\EndIf	
            	
			\EndFor
			\LState $D_{rep}^p=merge(D^p(G_1),D^p(G_2),\dots, D^p(G_n)) $  
            \If {rep==t}
            \LState $D^p=merge(D^p_1,D^p_2,\dots, D^p_t) $ 					\LState randomly swap the tuple of $D^p$ 
			\LState release $\ D^{p} $	
            \LState rep=0
            \EndIf
			\Until the data stream is stopped or paused
					
			\Statex \textbf{End Algorithm}
		\end{algorithmic}
	\end{spacing}
\end{algorithm}

\subsection{Variations of the proposed algorithm}
\label{configurations}
The proposed algorithm is configurable and allows the use of different settings. Four versions of the proposed algorithm were derived: two for static datasets (static setting) and two for data streams (dynamic setting). The version for static data achieves better results than existing schemes. The version for streams offers a solution for a new, emerging problem that has not received sufficient research interest so far. Brief descriptions of the algorithms are provided in Table \ref{alvariations}. The variations were evaluated and compared for their accuracy and security against privacy attacks (more details are available in Section \ref{results}). 

\begin{table}[htbp]
   \centering
   \caption{Variations/Derivations of the proposed algorithm}
   \resizebox{1\columnwidth}{!}{
     \begin{tabular}{|l|p{39.355em}|}
     \toprule
     \textbf{Algorithm} & \multicolumn{1}{l|}{\textbf{Description/Application}} \\
     \midrule
     $P^2RoCAl$ & The umbrella concept (\textbf{P}rivacy \textbf{P}reserving \textbf{Ro}tation based \textbf{C}ondensation \textbf{Al}gorithm) which represents all its derivatives, given in the following rows of this table. \\
     \midrule
     $k'-P^2RoCAl$ & $k'-P^2RoCAl$ algorithm was used for the static data perturbation where $k'$ represents the group size, and the grouping is done using Algorithm \ref{datclust}. This algorithm was tested with different settings under different experiments. Therefore, the corresponding settings of $k'-P^2RoCAl$ are provided under each experiment with different settings.\\
     \midrule
     $k-P^2RoCAl\_kmeans$ & $k-P^2RoCAl\_kmeans$ algorithm was used for the static data perturbation where grouping is done using the $k-means$ algorithm and $k$ represents the number of groups. During the experiments, $k$ was increased from 5 to 35 in successive steps of 5.\\
     \midrule
     $k'-P^2RoCAl\_streams$ & $k'-P^2RoCAl\_streams$ algorithm was used for the stream data perturbation where $k'$ represents the group size. The grouping is done using Algorithm \ref{datclust}. During the experiments, $k'$ was increased from 100 to 500 in successive steps of 100 and the buffer size $l$ was fixed to 1000. \\
     \midrule
     $k-P^2RoCAl\_kmeans\_streams$ & $k-P^2RoCAl\_kmeans\_streams$ algorithm was used for the data stream perturbation where grouping is done using the $k-means$ algorithm and $k$ represents the number of groups. During the experiments, $k$ was increased from 5 to 35 in successive steps of 5 and the buffer size $l$ was fixed to 1000. \\
     \bottomrule
     \end{tabular}%
     }
   \label{alvariations}%
\end{table}%

\section{Results}
\label{results}
This section first provides information about the experimental setup and the resources used in the experiments.  Then it describes the experimental results for all the four variations/derivations of $P^2RoCAl$ explained in Table \ref{alvariations}, and then we compare the results with rotation perturbation (RP) and data condensation (DC). RP and DC were selected for comparison because $P^2RoCAl$ inherits some properties of RP and DC. More specifically, the RP's distance (between the tuples) preservation property and DC's clustering/grouping and efficient data processing properties were effectively used in $P^2RoCAl$.
\subsection{Experimental Setup}
\label{expsetup}

All features were tested on a Windows 7 (Enterprise 64-bit, Build 7601) computer with an Intel(R) i7-4790 (4$^{th}$ generation) CPU (8 core, 3.60 GHz) and 8192 MB RAM. The scalability of the proposed algorithm was tested using a Linux (SUSE Enterprise Server 11 SP4) SGI UV3000 supercomputer, with 64 Intel Haswell 10-core processors, 25MB cache and 8TB of global shared memory connected by SGI's NUMAlink interconnect. The algorithm was implemented in MATLAB R2016b. Data classification tests were carried out by using Weka 3.6 \cite{witten2016data} which is a collection of machine learning algorithms for data mining tasks.

\subsubsection{Datasets used for testing and  comparison}
The datasets used for performance testing have different dimensions and vary from small to large, and contain only numerical attributes apart from the class attribute. A short description of the seven datasets used for testing is given in Table \ref{datasettb}. The efficiency of the algorithms for the data stream case ($k'-P^2RoCAl\_streams$ and $k-P^2RoCAl\_kmeans\_streams$ ) is very closely related to that of the static case, as we use a buffer size ($l$) in the case of data streams. To test the performance of $k'-P^2RoCAl\_streams$ and $k-P^2RoCAl\_kmeans\_streams$ for the data stream case, the same datasets which are shown in Table \ref{datasettb} were used with the dynamic settings of Algorithm \ref{datastreams}.

\begin{table}[htbp]
  \centering
  \caption{Short descriptions of the datasets selected for testing}
  \resizebox{0.55\columnwidth}{!}{
    \begin{tabular}{|p{10em}|l|r|r|r|}
    \toprule
    \multicolumn{1}{|l|}{\textbf{Dataset}} & \textbf{Abbreviation} & \multicolumn{1}{p{4.215em}|}{\textbf{Number\newline{}of\newline{}Records}} & \multicolumn{1}{p{5.645em}|}{\textbf{Number\newline{}of\newline{}Attributes}} & \multicolumn{1}{p{4.215em}|}{\textbf{Number\newline{}of\newline{}Classes}} \\
    \midrule
    \midrule
    \multicolumn{1}{|l|}{Winequality\tablefootnote{https://archive.ics.uci.edu/ml/datasets/Wholesale+customers}} & WQDS  & 4898  & 11    & 7 \\
    \midrule
    \multicolumn{1}{|l|}{Page Blocks\tablefootnote{https://archive.ics.uci.edu/ml/datasets/Page+Blocks+Classification}} & PBDS  & 5473  & 11    & 5 \\
    \midrule
    \multicolumn{1}{|l|}{Epileptic Seizure \tablefootnote{https://archive.ics.uci.edu/ml/datasets/Epileptic+Seizure+Recognition}} & ESDS  & 11500  & 179    & 5 \\
    
    \midrule
    \multicolumn{1}{|l|}{Fried\tablefootnote{https://www.openml.org/d/901}} & FRDS  & 40769 & 11    & 2 \\

    \midrule
    Statlog\tablefootnote{https://archive.ics.uci.edu/ml/datasets/Statlog+\%28Shuttle\%29} & SSDS  & 43501 & 10     & 5 \\
    \midrule
    Hepmass\tablefootnote{https://archive.ics.uci.edu/ml/datasets/HEPMASS} & HPDS  & 3310816 & 28     & 2 \\
    \midrule
    Higgs\tablefootnote{https://archive.ics.uci.edu/ml/datasets/HIGGS} & HIDS  & 11000000 & 28     & 2 \\
    \bottomrule
    \end{tabular}%
}
  \label{datasettb}%
\end{table}%

\subsubsection{Classification algorithms used for testing and comparison}
\label{classificationalgo}
First, the dynamics of classification accuracy of $P^2RoCAl$ were measured and compared with the results of data condensation (DC) using a k-nearest neighbor (kNN) \cite{witten2016data} classification approach (for k=1). As different classes of classification algorithms employ different classification strategies \cite{lessmann2015benchmarking}, the classification accuracy of our methods were tested and compared against four more classification algorithms, namely: decision table \cite{witten2016data}, naive Bayes \cite{witten2016data}, random tree \cite{witten2016data} and J48 \cite{quinlan1993c4}. Decision table builds a decision table majority classifier. Naive Bayes is a fast classification algorithm based on probabilistic classifiers. Random tree constructs a tree that considers $K$ randomly chosen attributes at each node while performing no pruning. J48 is an implementation of the decision tree based classification algorithm \cite{witten2016data}.

\subsection{Performance Evaluation of the Perturbation Algorithm}
Performance evaluation was focused on three factors: classification accuracy, attack resilience, and time consumption. The results were compared with the results obtained from random rotation perturbation (RP) \cite{chen2005random} and data condensation (DC) \cite{aggarwal2004condensation}. Seven datasets (Table \ref{datasettb}) were used to test and compare the proposed algorithm and its derivations (Table \ref{alvariations}). The datasets perturbed by $P^2RoCAl$ were tested for classification accuracy using the k-nearest neighbor (kNN) classification algorithm where the value of $k$ for kNN was maintained at a constant of 1 throughout all the experiments. kNN  is a non-parametric method used for classification \cite{witten2016data}. The original datasets were perturbed using RP and DC also, and the classification accuracy results were compared. The comparisons were conducted using the nonparametric statistical comparison test, Friedman's rank test, which is analogous to a standard one-way repeated-measures analysis of variance \cite{howell2016fundamental}. Friedman's mean ranks (FMR) and the statistical significance of the results were recorded. 

The resilience of the method was tested against three attack types, to which the proposed method is most vulnerable. Section \ref{attacktesting} provides a detailed description of the three attack types. $P^2RoCAl$'s attack resilience results were compared with RP and DC using Friedman's rank test, and the results were presented with the corresponding test statistics. 

Runtime complexity of  $P^2RoCAl$ was evaluated, and then time consumption experiments were run on the ESDS dataset to check the effect of the number of tuples and number of attributes on the time consumption. The ESDS dataset was specially selected for this analysis as it has a high number of attributes (Table \ref{datasettb}). Next, the running times of DC, RP, $k'-P^2RoCAl$ and $k-P^2RoCAl\_kmeans$ were measured for the four datasets, PBDS, WQDS, FRDS, and FRDS. The results were then compared with each other using Friedman's rank test. 

\subsubsection{Classification Accuracy} 
\begin{figure}[H]
	\centering
	\subfloat[PBDS dataset]{\includegraphics[width=0.49\textwidth, trim=0cm 0cm 0cm 0cm]{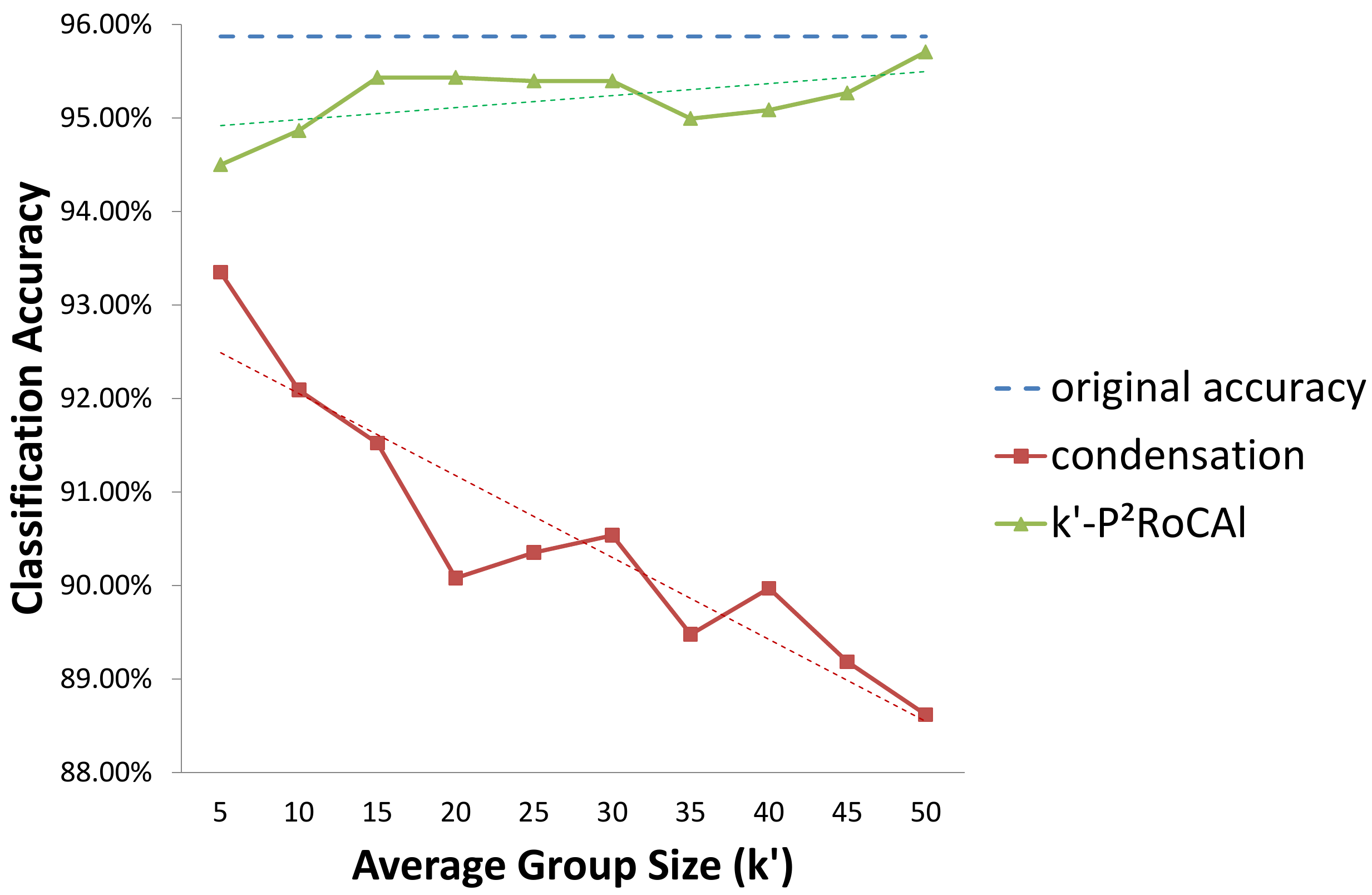}\label{pbdsaccuracy}}
		\hfill
	\subfloat[WQDS dataset]{\includegraphics[width=0.49\textwidth, trim=0.3cm 0cm 0cm 0cm]{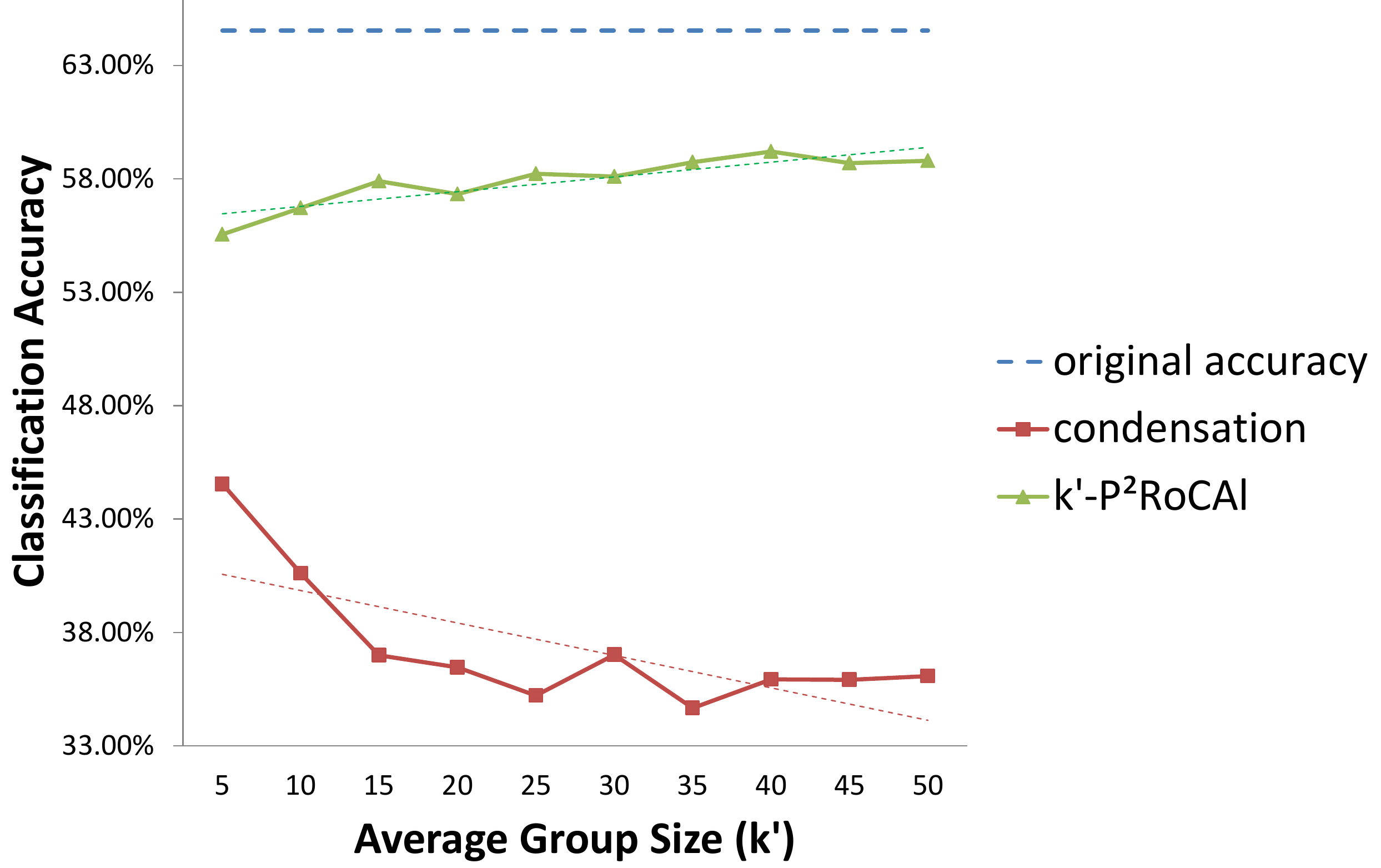}\label{wqdsaccuracy}}
	\hfill	
	\subfloat[FRDS dataset]{\includegraphics[width=0.49\textwidth, trim=0cm 0cm 0cm 0cm]{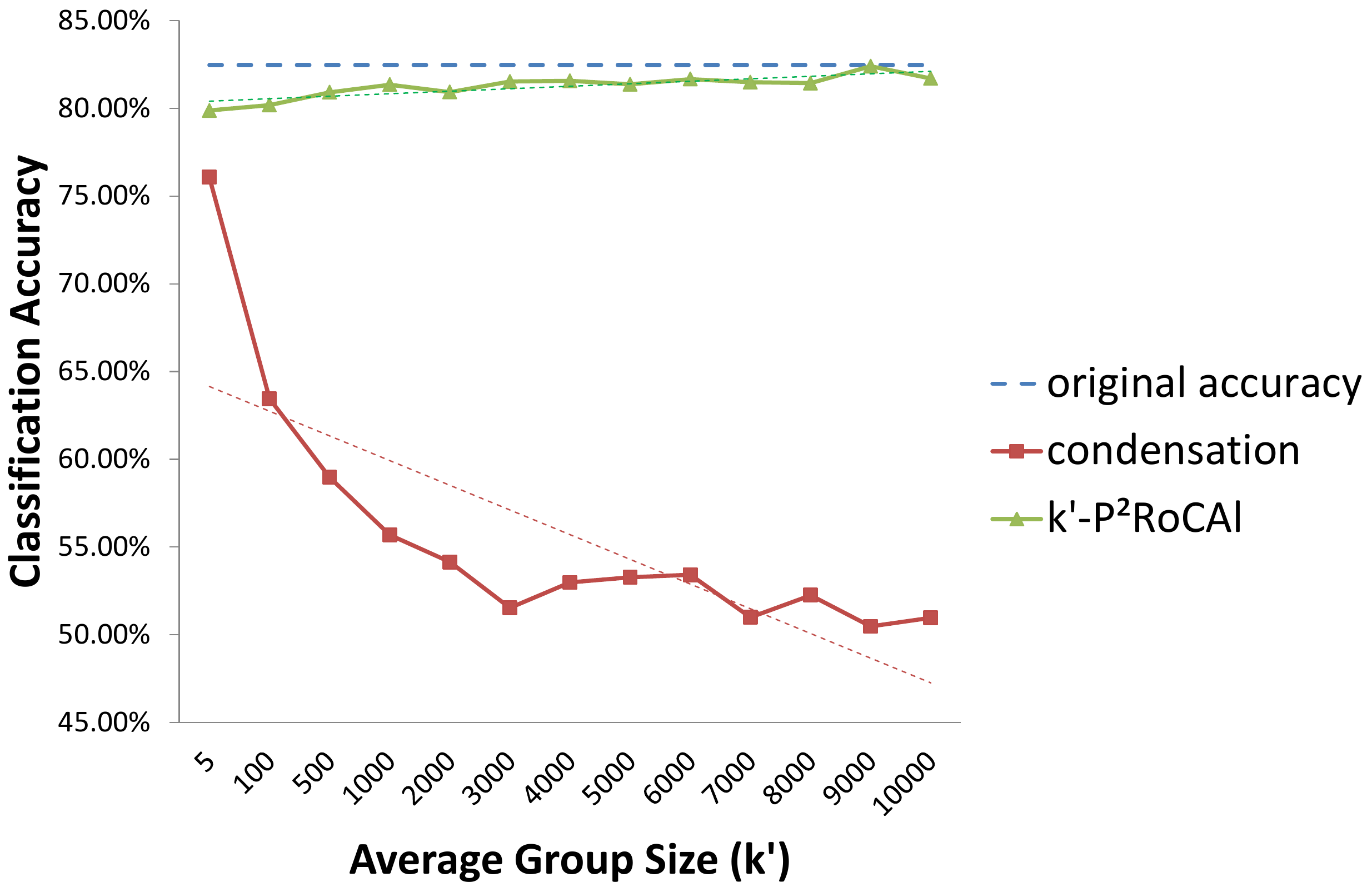}\label{afdsaccuracy}}
	\hfill
	\subfloat[SSDS dataset]{\includegraphics[width=0.49\textwidth, trim=0.3cm 0cm 0cm 0cm]{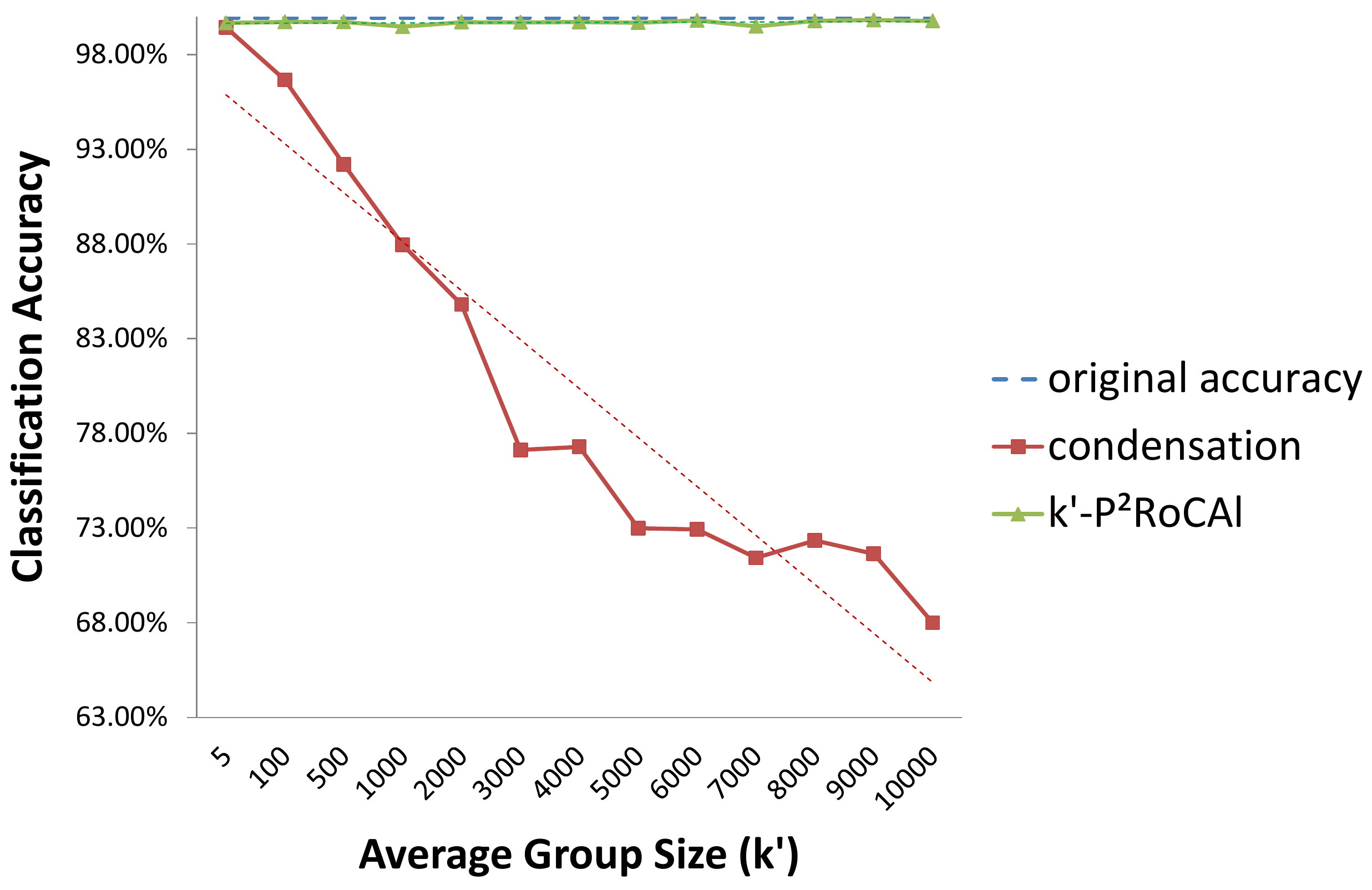}\label{ssdsaccuracy}}	
	\caption{Classification accuracy comparison of the perturbed data for large k$'$ values}\label{classificationaccuracy}
    \label{classaccuracy}
\end{figure}

Figure \ref{classificationaccuracy} shows the average classification accuracy returned by the kNN classification algorithm against increasing  $k'$ (the group size) sizes for $k'-P^2RoCAl$ and DC. Table \ref{classyaccuracy} has the average classification accuracy values returned by kNN classification algorithm for the original datasets (PBDS, WQDS, FRDS and SSDS) and the datasets perturbed by $k'-PR^2oCAl$, RP and DC. The results were produced with 10-fold cross validation. 

\begin{figure}[H]
	\centering
	\scalebox{0.7}{
		\includegraphics[width=1\textwidth, trim=0cm 0cm 0cm
		0cm]{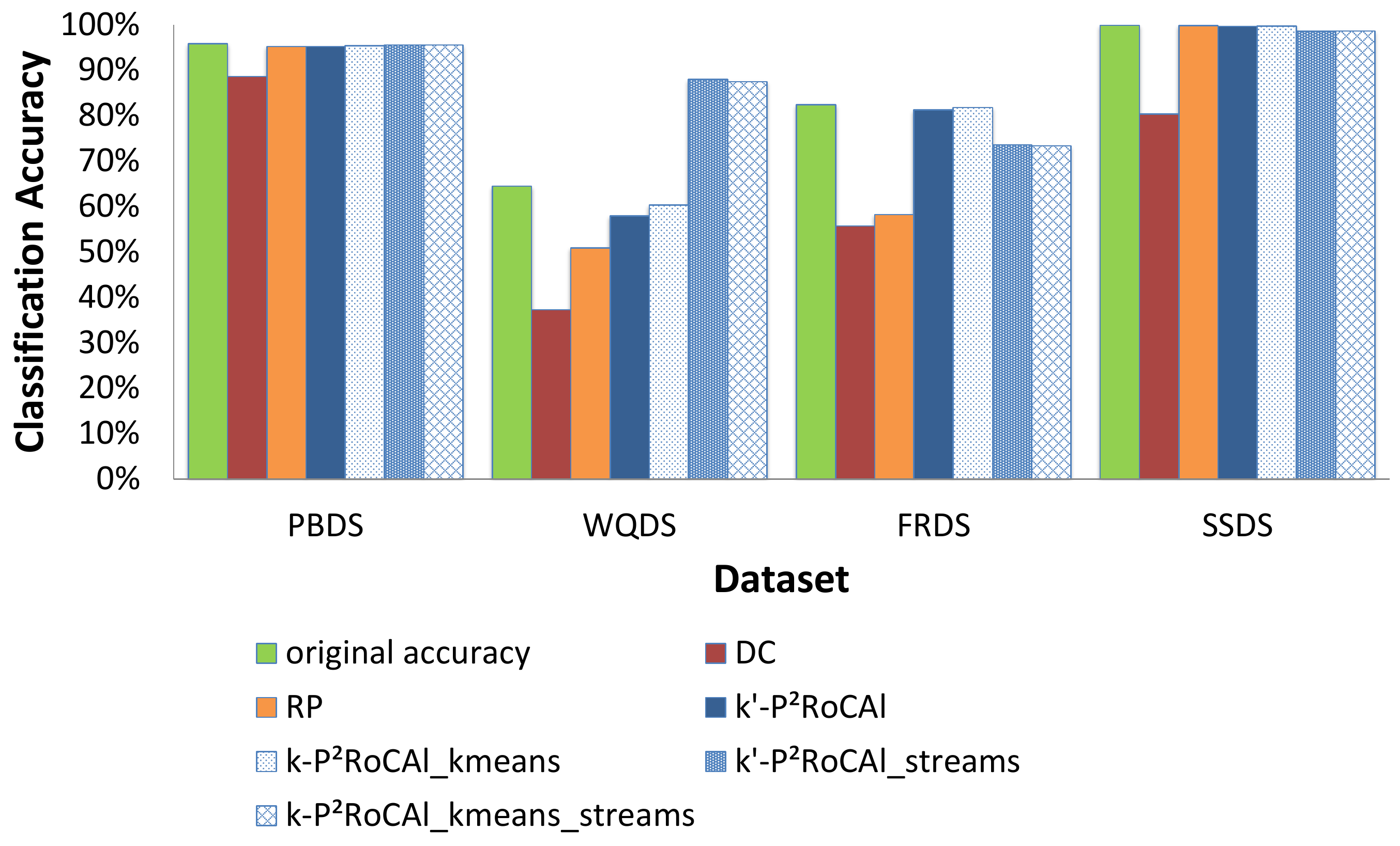}
	}
	\caption{Average accuracy provided by the methods}
	\label{averageaccuracy}
\end{figure}

The last column in Table \ref{classyaccuracy} has the Friedman's test ranks (denoted by FMR) returned for each of the methods. The test statistics of the experiment had a $\chi^2$ value of 14.786, degree of freedom of 6 and a p-value of 0.022. The p-value suggests that there is a significant difference between the average accuracies of the datasets. The original dataset has the highest mean rank, as it has the highest classification accuracy. The mean ranks prove that $P^2RoCAl$ and its variations provide the highest classification accuracy. Using kNN classification on the four datasets (as given in Table \ref{classyaccuracy}), DC returns the least accurate results, and $k-P^2RoCAl\_kmeans$ and $k'-P^2RoCAl\_streams$ produce the best results. The results available in Table \ref{classyaccuracy}, are further illustrated using a bar graph in Figure \ref{averageaccuracy}.  

\begin{table}[htbp]
  \centering
  \caption{Average accuracy returned by the methods on the datasets}
  \resizebox{1\columnwidth}{!}{
    \begin{tabular}{|l|r|r|r|r|r|r|r|}
    \toprule
    \textbf{Dataset} & \multicolumn{1}{p{7.715em}|}{\textbf{original \newline{}accuracy}} & \multicolumn{1}{l|}{\textbf{DC}} & \multicolumn{1}{p{6.43em}|}{\textbf{RP}} & \multicolumn{1}{l|}{\textbf{$k'-P^2RoCAl$}} & \multicolumn{1}{l|}{\textbf{$k-P^2RoCAl\_kmeans$}} & \multicolumn{1}{l|}{\textbf{$k'-P^2RoCAl\_streams$}} & \multicolumn{1}{l|}{\textbf{$k-P^2RoCAl\_kmeans\_streams$}} \\
    \midrule
    PBDS  & 95.87\% & 88.62\% & 95.27\% & 95.23\% & 95.45\% & 95.60\% & 95.58\% \\
    \midrule
    WQDS  & 64.54\% & 37.34\% & 50.88\% & 58\%  & 60.35\% & 88.01\% & 87\% \\
    \midrule
    FRDS  & 82.46\% & 55.71\% & 58.26\% & 81.26\% & 81.84\% & 73.57\% & 73\% \\
    \midrule
    SSDS  & 99.94\% & 80.37\% & 99.85\% & 99.70\% & 99.79\% & 98.59\% & 99\% \\
    \midrule
    \midrule
    \textbf{FMR} & 6.5   & 1     & 3.25  & 3.5   & 4.75  & 4.75  & 4.25 \\
    \bottomrule
    \end{tabular}%
    }
  \label{classyaccuracy}%
\end{table}%

Experiments were conducted to investigate the dynamics of classification accuracy for data produced by the three methods: RP, DC and $k'-P^2RoCAl$. Table \ref{classaccuracycomp} contains the classification accuracies of the data produced by the three methods along with the original data sets for four classification algorithms: Decision Table, J48, Naive Bayes and Random Tree. 

\begin{table}[htbp]
	\centering
	\caption{Average classification accuracies on different classification algorithms}
	\resizebox{0.8\columnwidth}{!}{
	\begin{tabular}{|c|l|r|r|r|r|}
		\toprule
	\multicolumn{1}{|l|}{\textbf{Dataset}} 	&   \multicolumn{1}{l|}{\textbf{Algorithm}}     & \multicolumn{1}{l|}{\textbf{Decision Table}} & \multicolumn{1}{l|}{\textbf{J48}} & \multicolumn{1}{l|}{\textbf{Naive Bayes}} & \multicolumn{1}{l|}{\textbf{Random Tree}} \\
		\midrule
		\multirow{4}[8]{*}{\textbf{WQDS}} & \textbf{Original} & 53.27\% & 59.82\% & 44.67\% & 61.70\% \\
		\cmidrule{2-6}          & \textbf{RP} & 45.75\% & 46.47\% & 33.83\% & 46.71\% \\
		\cmidrule{2-6}          & \textbf{DC} & 45.57\% & 39.93\% & 37.85\% & 36.79\% \\
		\cmidrule{2-6}          & \textbf{$k'-P^2RoCAl$} & 44.99\% & 53.00\% & 40.55\% & 57.61\% \\
		\midrule
		\multirow{4}[8]{*}{\textbf{PBDS}} & \textbf{Original} & 95.63\% & 96.88\% & 90.85\% & 96.05\% \\
		\cmidrule{2-6}          & \textbf{RP} & 93.88\% & 95.16\% & 78.46\% & 94.13\% \\
		\cmidrule{2-6}          & \textbf{DC} & 93.21\% & 92.48\% & 71.43\% & 90.30\% \\
		\cmidrule{2-6}          & \textbf{$k'-P^2RoCAl$} & 93.48\% & 94.46\% & 39.11\% & 93.90\% \\
		\midrule
		\multirow{4}[8]{*}{\textbf{FRDS}} & \textbf{Original} & 83.42\% & 89.41\% & 86.53\% & 86.95\% \\
		\cmidrule{2-6}          & \textbf{RP} & 63.27\% & 63.32\% & 66.12\% & 58.25\% \\
		\cmidrule{2-6}          & \textbf{DC} & 61.09\% & 61.30\% & 60.97\% & 56.15\% \\
		\cmidrule{2-6}          & \textbf{$k'-P^2RoCAl$} & 72.58\% & 77.83\% & 65.71\% & 76.25\% \\
		\midrule
		\multirow{4}[8]{*}{\textbf{SSDS}} & \textbf{Original} & 99.72\% & 99.96\% & 91.84\% & 99.96\% \\
		\cmidrule{2-6}          & \textbf{RP} & 98.36\% & 99.65\% & 78.54\% & 99.60\% \\
		\cmidrule{2-6}          & \textbf{DC} & 83.78\% & 85.38\% & 74.09\% & 80.04\% \\
		\cmidrule{2-6}          & \textbf{$k'-P^2RoCAl$} & 95.40\% & 99.49\% & 75.26\% & 99.51\% \\
		\bottomrule
	\end{tabular}%
}
	\label{classaccuracycomp}%
\end{table}%

\begin{figure}[H]
	\centering
	\scalebox{0.7}{
		\includegraphics[width=1\textwidth, trim=0cm 0cm 0cm
		0cm]{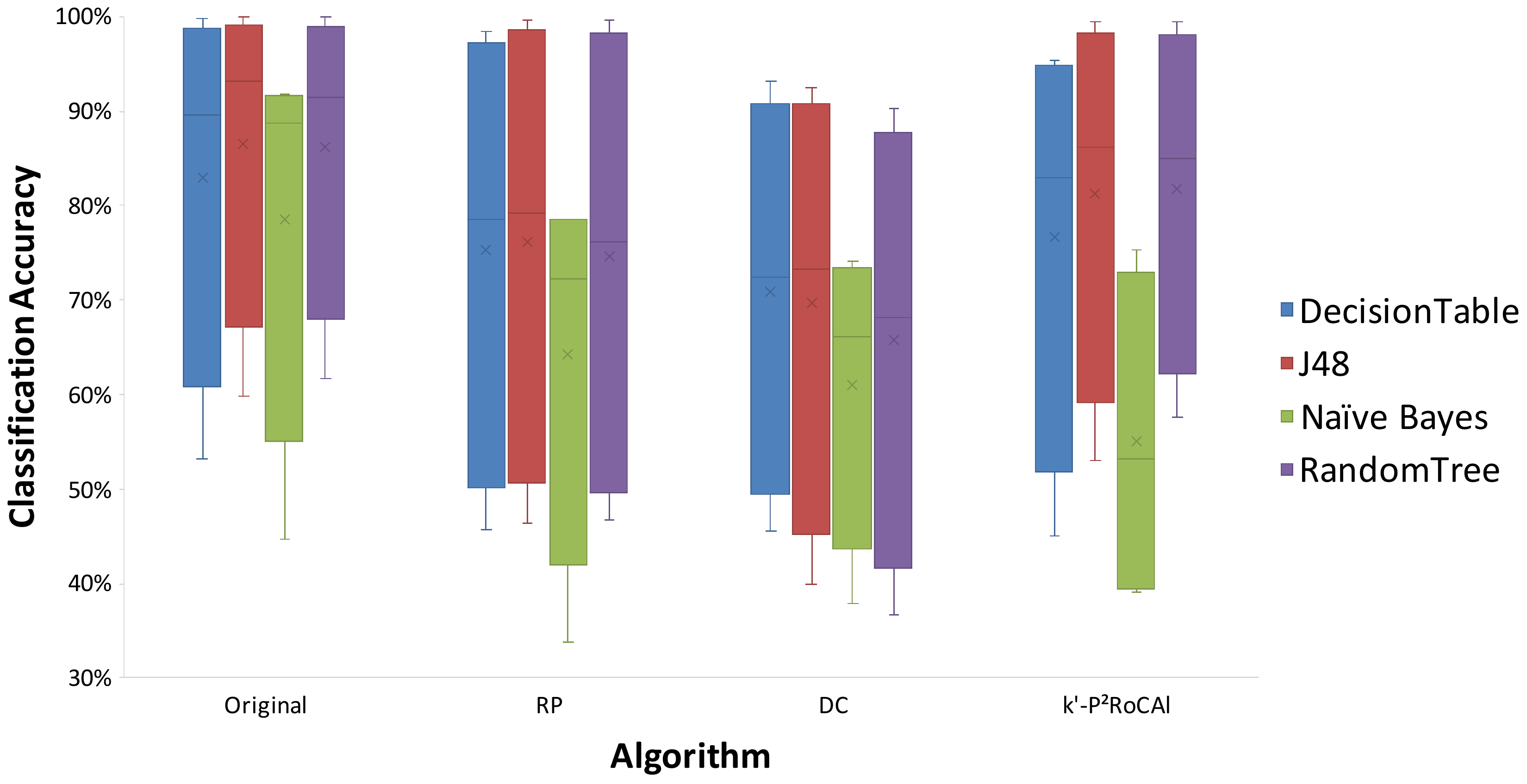}
	}
	\caption{Box plots on the data available in Table \ref{classaccuracycomp}}
	\label{accuracybox}
\end{figure}

Figure \ref{accuracybox} has the box plots for the data in Table \ref{classaccuracycomp}. It can be noted that the classification accuracy result for $P^2RoCAl$ is better than the result for DC and very close to the results for RP.

\subsubsection{Attack Resistance}
\label{attacktesting}
\begin{figure}[H]
	\centering
	\subfloat[PBDS dataset minimum STD(D-D$^p$)]{\includegraphics[width=0.49\textwidth, trim=0cm 0cm 0cm 0cm]{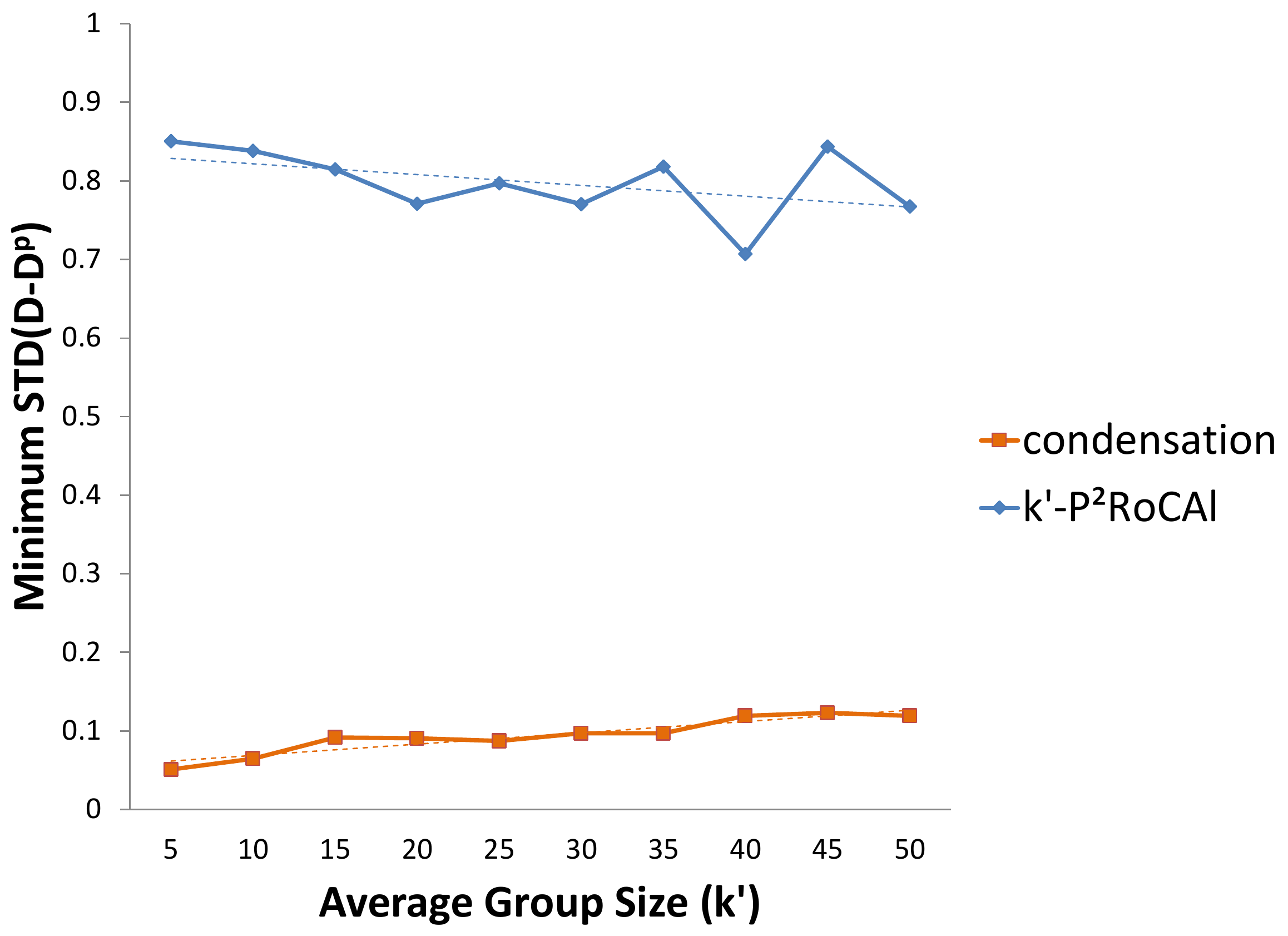}\label{pbdsstdmin}}
	\hfill
	\subfloat[PBDS dataset average STD(D-D$^p$)]{\includegraphics[width=0.49\textwidth, trim=0.3cm 0cm 0cm 0cm]{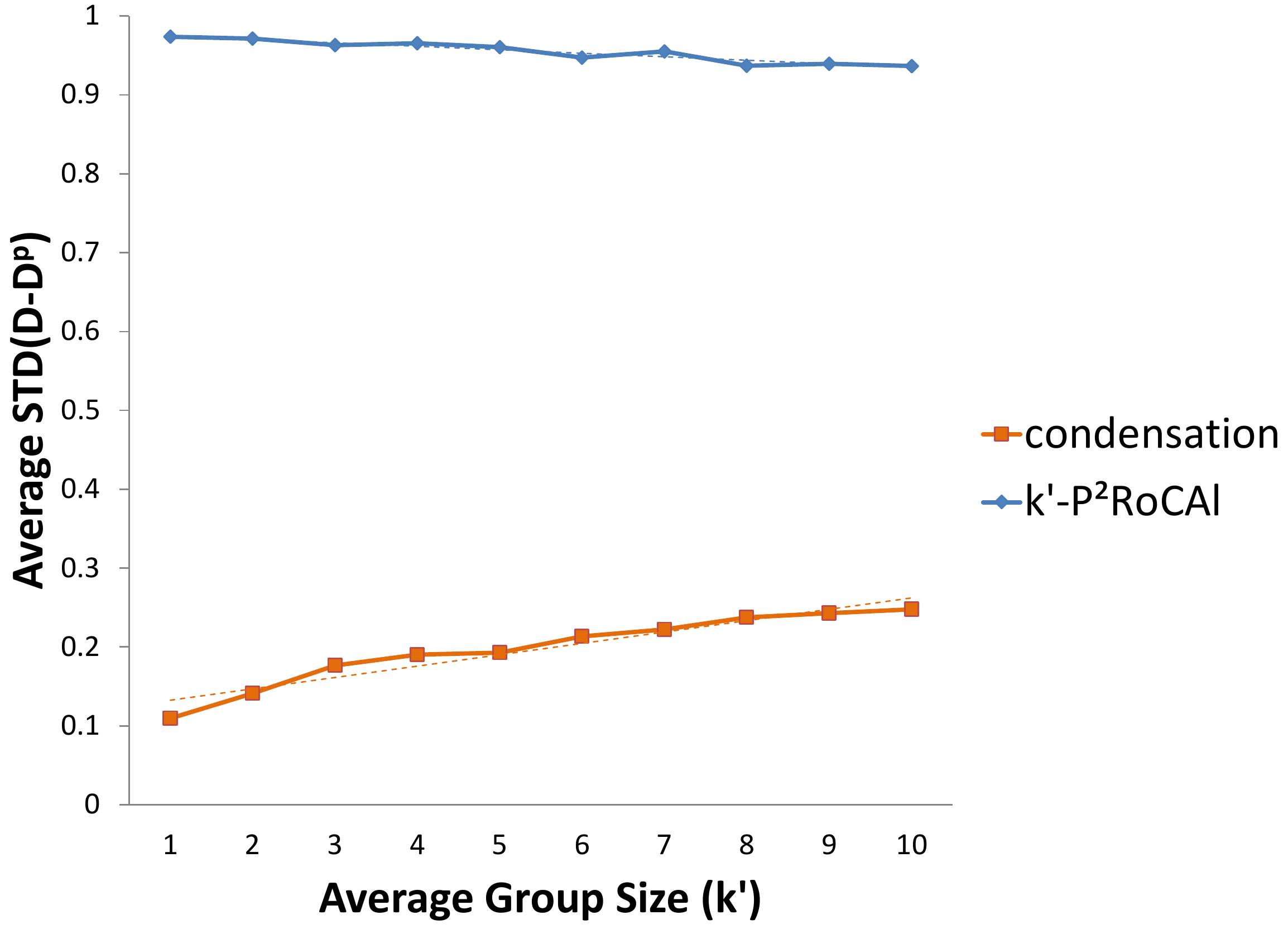}\label{pbdsstdavg}}
	\hfill	
	\subfloat[SSDS dataset minimum STD(D-D$^p$)]{\includegraphics[width=0.49\textwidth, trim=0cm 0cm 0cm 0cm]{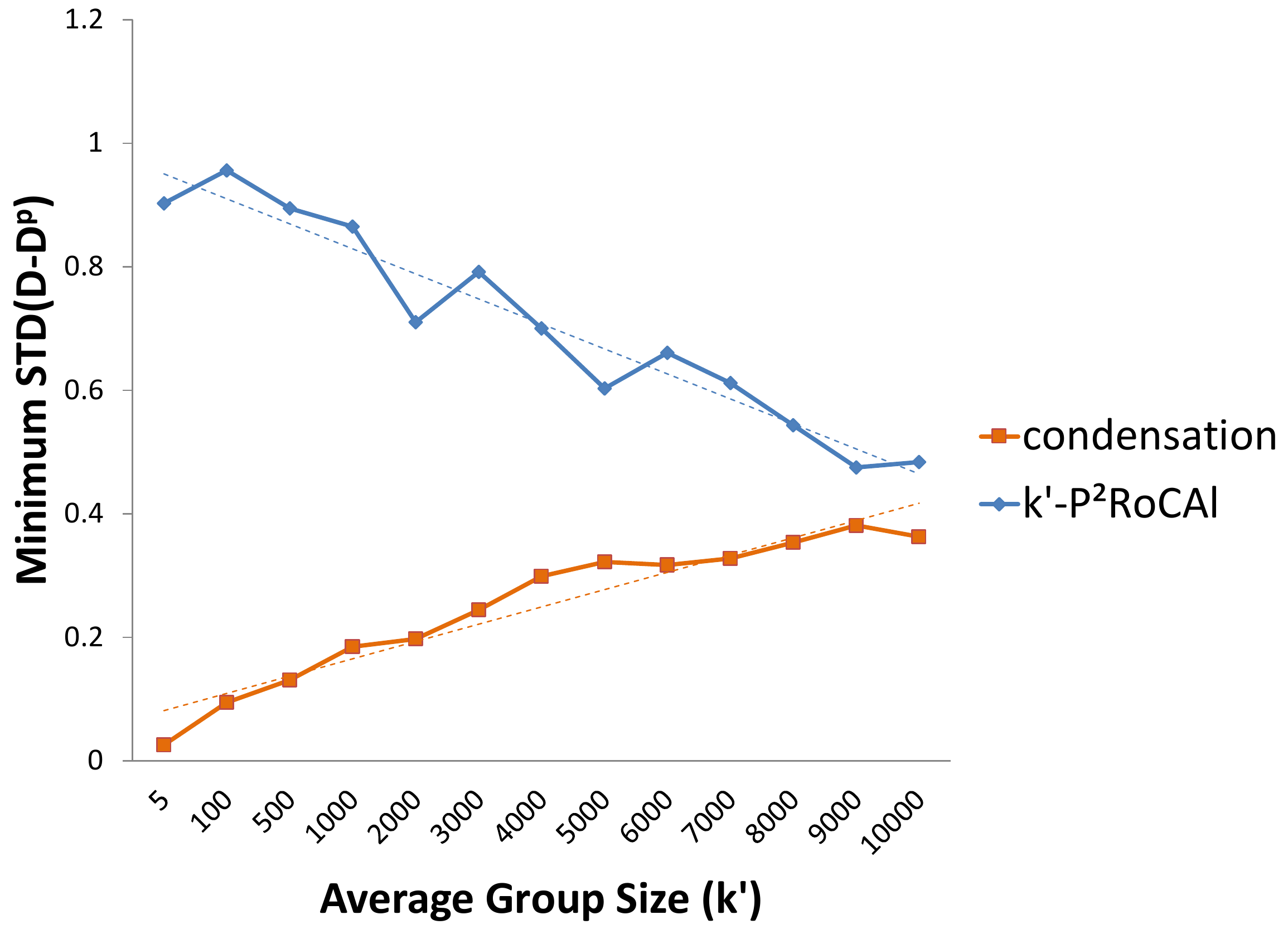}\label{ssdsstdmin}}
	\hfill
	\subfloat[SSDS dataset average STD(D-D$^p$)]{\includegraphics[width=0.49\textwidth, trim=0.3cm 0cm 0cm 0cm]{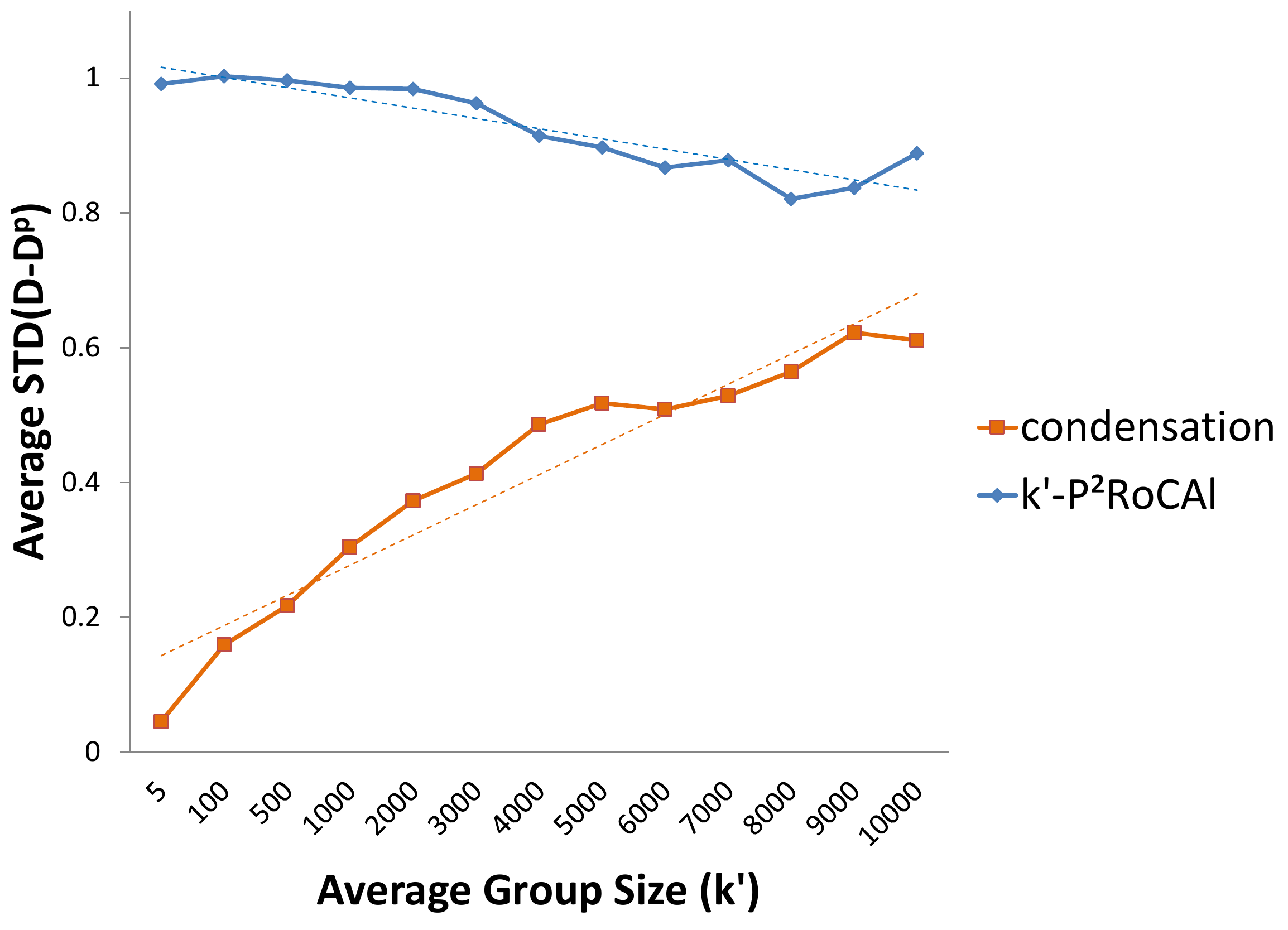}\label{ssdsstdavg}}
	\caption{Dynamics of STD(D-D$^p$) against increasing  k$'$ values}\label{stddifferenceagainstk}
\end{figure}

The literature shows different probable attack types against matrix multiplicative data perturbation \cite{okkalioglu2015survey}. The main purpose of these attacks is to restore the original data from the perturbed data. The larger the difference between original data and perturbed data, the more difficult the attack becomes. Figure \ref{stddifferenceagainstk} shows the variation of $STD(D-D^p)$ against increasing $k'$ for $P^2RoCAl$ and DC. This captures the variation of the standard deviation of the difference between the original and perturbed data, and indicates how secure the methods are when using different group sizes ($k'$).

$P^2RoCAl$ is based on matrix multiplication, and it was tested against three types of attacks: naive estimation,  accessible/known I/O (assuming 10\% of the original data is known to the adversary) and ICA based attacks. To check the resilience of the proposed algorithm against ICA based attacks, the procedure described in \cite{chen2005random} was employed, and the FastICA package \cite{gavert2005fastica} was used to evaluate the effectiveness of ICA-based reconstruction of the perturbed data. The attack resilience experiments were inspired by those described in \cite{chen2005random,chen2011geometric}, and the results are presented in Table \ref{attackresilience}. For naive inference attacks, the standard deviation of the difference between the original data and the perturbed data (both normalized) was calculated. For ICA and IO, the standard deviation of the difference between the original and reconstructed data (both normalized) was used. The results returned by RP were generated using 10 number of iterations with a noise factor (sigma) of 0.3 (the default setting). The minimum (``$min$") values under each test indicate the minimum guarantee of resilience while the average (``$avg$") values give an impression of overall resilience.

\begin{table}[H]
	\centering
	
	\caption{Results of the attack resilience experiments} 
	\label{attackresilience}
	
	\begin{small}
		\resizebox{1\columnwidth}{!}{
	\begin{tabular}{|c|l|r|r|r|r|r|r|}
		\toprule
		\multicolumn{1}{|l|}{\textbf{Dataset}} & \textbf{Algorithm} & \multicolumn{1}{l|}{\textbf{NImin}} & \multicolumn{1}{l|}{\textbf{NIavg}} & \multicolumn{1}{l|}{\textbf{ICAmin}} & \multicolumn{1}{l|}{\textbf{ICAavg}} & \multicolumn{1}{l|}{\textbf{IOmin}} & \multicolumn{1}{l|}{\textbf{IOavg}} \\
		\midrule
		\multirow{6}[12]{*}{PBDS} & DC & 0.0940 & 0.1975 & 0.5595 & 0.6937 & 0.2507 & 0.5278 \\
		\cmidrule{2-8}          & RP    & 0.1022 & 0.3214 & 0.5883 & 0.7444 & 0.0002 & 0.1446 \\
		\cmidrule{2-8}          & $k'-P^2RoCAl$ & 0.7977 & 0.9550 & 0.7009 & 0.7075 & 0.6892 & 0.7046 \\
		\cmidrule{2-8}          & $k-P^2RoCAl\_kmeans$ & 0.3901 & 0.6182 & 0.7006 & 0.7071 & 0.6961 & 0.7052 \\
		\cmidrule{2-8}          & $k'-P^2RoCAl\_streams$ & 0.5436 & 0.7633 & 0.6873 & 0.7067 & 0.6702 & 0.7008 \\
		\cmidrule{2-8}          & $k-P^2RoCAl\_kmeans\_streams$ & 0.6509 & 0.8052 & 0.6938 & 0.7076 & 0.6848 & 0.7016 \\
		\midrule
		\multirow{6}[12]{*}{WQDS} & DC & 0.0278 & 0.0652 & 0.5640 & 0.6938 & 0.1077 & 0.5762 \\
		\cmidrule{2-8}          & RP    & 0.0338 & 0.1090 & 0.6272 & 0.6840 & 0.0057 & 0.4063 \\
		\cmidrule{2-8}          & $k'-P^2RoCAl$ & 0.4030 & 0.8199 & 0.6993 & 0.7072 & 0.6914 & 0.7038 \\
		\cmidrule{2-8}          & $k-P^2RoCAl\_kmeans$ & 0.4030 & 0.8199 & 0.6993 & 0.7072 & 0.6914 & 0.7038 \\
		\cmidrule{2-8}          & $k'-P^2RoCAl\_streams$ & 0.4083 & 0.7928 & 0.6734 & 0.7074 & 0.6319 & 0.6945 \\
		\cmidrule{2-8}          & $k-P^2RoCAl\_kmeans\_streams$ & 0.4808 & 0.8921 & 0.6867 & 0.7038 & 0.6598 & 0.6972 \\
		\midrule
		\multirow{6}[12]{*}{SSDS} & DC & 0.2493 & 0.4116 & 0.6374 & 0.7243 & 0.3895 & 0.5709 \\
		\cmidrule{2-8}          & RP    & 0.1994 & 0.3904 & 0.4781 & 0.7116 & 0.0019 & 0.0357 \\
		\cmidrule{2-8}          & $k'-P^2RoCAl$ & 0.7076 & 0.9250 & 0.7047 & 0.7070 & 0.7023 & 0.7062 \\
		\cmidrule{2-8}          & $k-P^2RoCAl\_kmeans$ & 0.4961 & 0.7221 & 0.7040 & 0.7067 & 0.7032 & 0.7063 \\
		\cmidrule{2-8}          & $k'-P^2RoCAl\_streams$ & 0.8386 & 0.9615 & 0.7035 & 0.7065 & 0.7017 & 0.7056 \\
		\cmidrule{2-8}          & $k-P^2RoCAl\_kmeans\_streams$ & 0.8391 & 0.9837 & 0.7054 & 0.7071 & 0.7006 & 0.7053 \\
		\midrule
		\multirow{6}[12]{*}{FRDS} & DC & 0.9434 & 1.0966 & 0.6181 & 0.7041 & 0.5686 & 0.6167 \\
		\cmidrule{2-8}          & RP    & 0.9309 & 1.0093 & 0.4847 & 0.6948 & 0.3956 & 0.5076 \\
		\cmidrule{2-8}          & $k'-P^2RoCAl$ & 1.1761 & 1.2963 & 0.7049 & 0.7072 & 0.7016 & 0.7063 \\
		\cmidrule{2-8}          & $k-P^2RoCAl\_kmeans$ & 1.1895 & 1.3131 & 0.7051 & 0.7074 & 0.7015 & 0.7062 \\
		\cmidrule{2-8}          & $k'-P^2RoCAl\_streams$ & 1.2897 & 1.3375 & 0.7043 & 0.7071 & 0.7011 & 0.7059 \\
		\cmidrule{2-8}          & $k-P^2RoCAl\_kmeans\_streams$ & 1.3153 & 1.3368 & 0.7046 & 0.7072 & 0.7012 & 0.7062 \\

		\bottomrule
	\end{tabular}%
}

\end{small} 
\end{table}%

The mean ranks produced by Friedman's rank test on the data available in Table \ref{attackresilience} are presented in Table \ref{friedmanattackranks}, with the test statistics: a $\chi^2$ value of 47.5694, a degree of freedom of 5 and a p-value of 4.3485e-09.  Table \ref{attackresilience} includes the standard deviation of the difference between original data and reconstructed data, except for naive inference, where the difference between original data and perturbed data is used. Here, a higher rank indicates better resilience. According to the p-value, the attack resilience rank averages of the six methods are significantly different. The mean ranks suggest that all the variations of the new method provide comparatively higher security against the privacy attacks and that DC provides the lowest level of resilience against the investigated attack methods. 

\begin{table}[htbp]
	\centering
	\caption{Friedman test ranks on the attack resilience data}
	\begin{tabular}{|l|r|}
		\toprule
		
		Method & \multicolumn{1}{l|}{Mean Rank} \\
		\midrule
		DC & 10.7917 \\
		\midrule
		RP    & 9.3333 \\
		\midrule
		$k'-P^2RoCAl$ & 24.4167 \\
		\midrule
		$k-P^2RoCAl\_kmeans$ & 21.5417 \\
		\midrule
		$k'-P^2RoCAl\_streams$ & 21.7917 \\
		\midrule
		$k-P^2RoCAl\_Kmeans\_streams$ & 23.1250 \\
		\bottomrule
	\end{tabular}%
	\label{friedmanattackranks}%
\end{table}%

\subsubsection{Time complexity}
\label{asymptotic}

\begin{figure}[h!]

	\subfloat[$k'-P^2RoCAl$ changing the number of tuples]{\includegraphics[width=0.49\textwidth, trim=0cm 0cm 0cm 0cm]{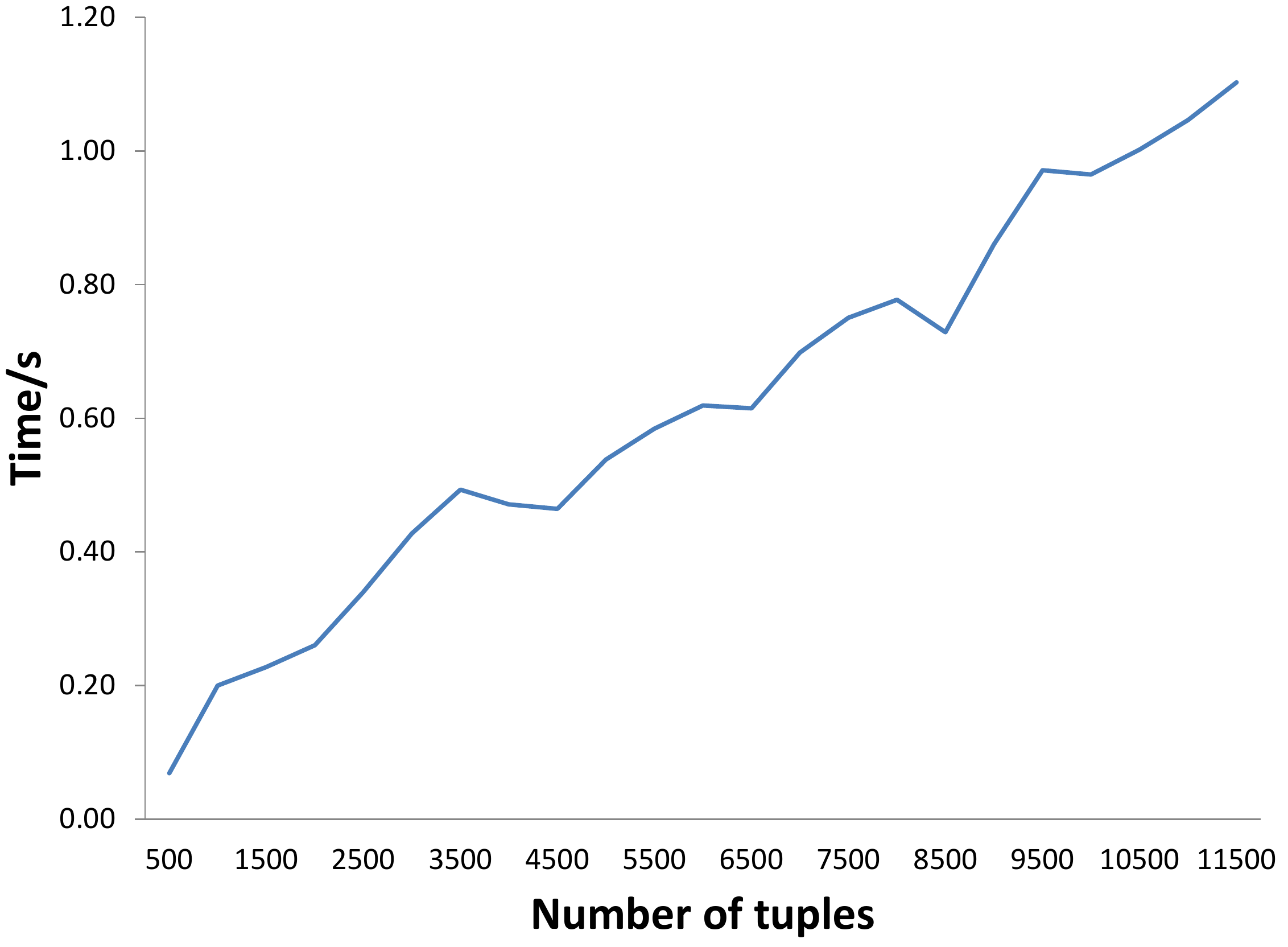}\label{kdashtuples}}
	\hfill
	\subfloat[$k'-P^2RoCAl$ changing the number of  attributes]{\includegraphics[width=0.49\textwidth, trim=0.3cm 0cm 0cm 0cm]{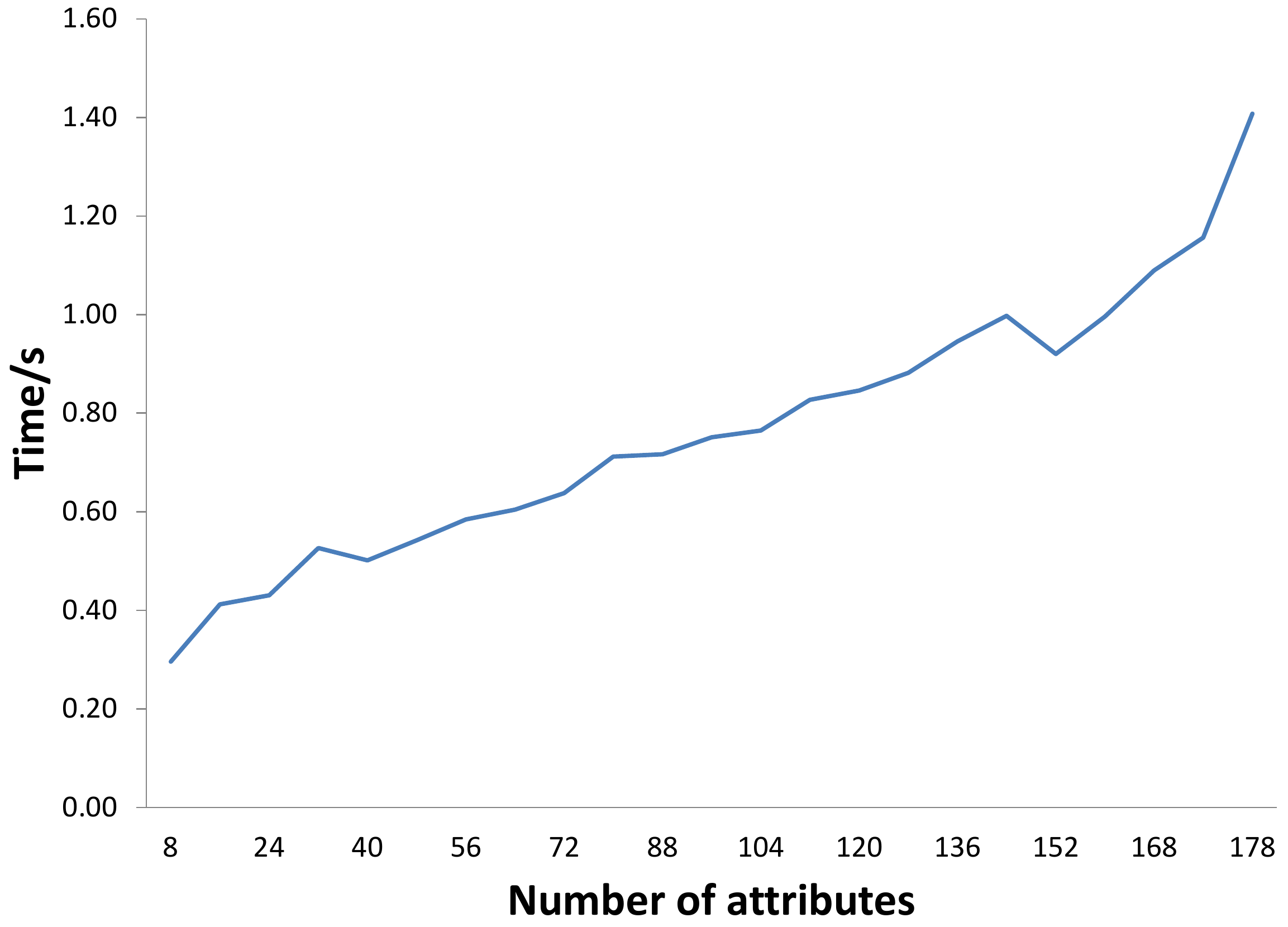}\label{kdashattribute}}
	\hfill	
	\subfloat[$k-P^2RoCAl\_kmeans$ changing the number of tuples]{\includegraphics[width=0.49\textwidth, trim=0cm 0cm 0cm 0cm]{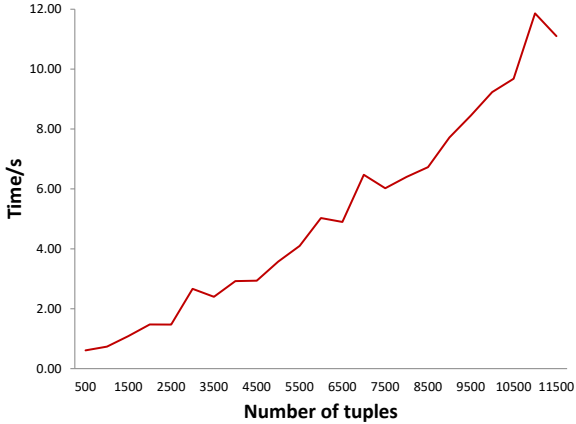}\label{kmeantuples}}
	\hfill
	\subfloat[$k-P^2RoCAl\_kmeans$ changing the number of attributes]{\includegraphics[width=0.49\textwidth, trim=0.3cm 0cm 0cm 0cm]{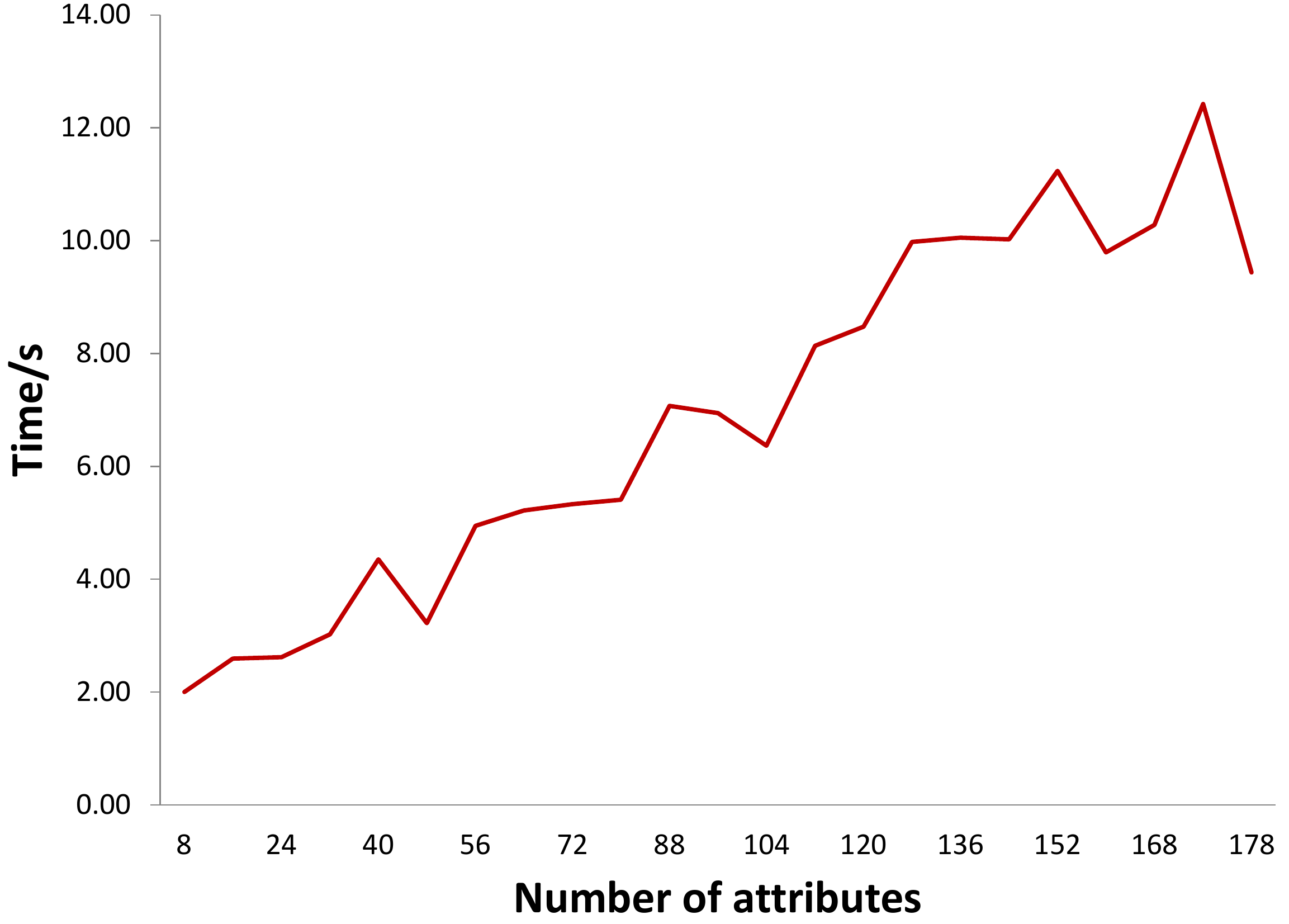}\label{kmeansattributes}}
	\caption{Time consumption patterns of $P^2RoCAl$}\label{timecomplexityplots}	
\end{figure}

Algorithm \ref{staticdata} has three main components that affect computational complexity: (i) clustering/grouping process (line \ref{line1}, Algorithm \ref{staticdata}), (ii) generating a rotation matrix for the rotation perturbation (line \ref{line5},  Algorithm \ref{staticdata}), and (iii) rotation perturbation on the groups (line \ref{line7} or \ref{line10}, Algorithm \ref{staticdata}). The first component has two ways to achieve the grouping, which is demonstrated in Algorithm \ref{datclust}. If the grouping is done using the k-means algorithm, the complexity would be $O(k\times m \times n \times I)$ where $I$ is the number of iterations. If the user chooses $k'$ as the option, then the clustering complexity would be $O(k'\times m \times (m / k') \times n)= O(m^2\times n) $  as the algorithm tries to select the closest $k'-1$ tuples per each randomly sampled tuple ($X_i$) from the dataset $D$ for $m/k'$ number of iterations.  For the second component, the algorithm first needs to conduct the eigenvalue decomposition, which has a computation complexity of $O(n^3)$. Next, the algorithm conducts the rotation perturbation, which has a computational complexity of  $O(m\times n^3)$. Figure \ref{timecomplexityplots} shows the time consumption of $k'-P^2RoCAl$ and $k-P^2RoCAl\_kmeans$ against changing the number of tuples and attributes of the ESDS dataset; when one was changing, the other was kept constant. During the time consumption analysis, $k'$ was maintained as constant at 1000 while $k$ was set to 10. As shown in Figure \ref{kdashtuples} and \ref{kmeantuples}, when the number of tuples is increased while the number of attributes is kept constant, the runtime complexity is governed by clustering. Hence the complexity is $O(k\times m \times I)$ if the user's choice is $k$, while the complexity is $O(m^2)$ when the user's choice of $k'$. When the number of tuples is maintained as constant, the worst case complexity of the algorithm becomes $O(n^3)$, but the plots Figure \ref{kdashattribute} and \ref{kmeansattributes} still show a pattern close to linear. This suggests that the effect of attributes is considerably low since $m>>>n$. Hence, the overall complexity is mainly governed by the number of tuples.

Table \ref{runtimeanalysis} shows the time consumption of DC, RP, $k'-P^2RoCAl$  and $k-P^2RoCAl\_kmeans$ for the perturbation of the four datasets: PBDS, WQDS, FRDS, and SSDS. The mean ranks produced by Friedman's rank test on the data in Table \ref{runtimeanalysis} are presented in the last row. The experiment returns the test statistics: a ($\chi^2$ value) of 8.4, a degree of freedom of 3 and a p-value of 0.038. Here, lower ranks suggest that the time consumption of the corresponding method is low compared to other methods.

\begin{table}[H]
	\centering
	\caption{Runtime analysis of the perturbation process consumed by the algorithms}
		\begin{tabular}{|l|r|r|r|r|}
			\toprule
			\textbf{Dataset} & {\textbf{DC/s}} & {\textbf{RP/s}} & {\textbf{$k'-P^2RoCAl$/s}} & {\textbf{$k-P^2RoCAl\_kmeans$/s}} \\
			\midrule
			PBDS  & 0.3421 & 1590.7000 & 0.3800 & 0.7462 \\
			\midrule
			WQDS  & 0.3022 & 217.0409 & 0.3439 & 0.9744 \\
			\midrule
			FRDS  & 2.4543 & 1807.6000 & 2.5760 & 1.1300 \\
			\midrule
			SSDS  & 2.7134 & 1269.6000 & 2.6830 & 4.2508 \\
			\midrule
            \midrule
			\textbf{FMR}   &  1.50     &   4.00    &   2.00    & 2.50 \\
			\bottomrule
		\end{tabular}%
	
	\label{runtimeanalysis}%
\end{table}%

According to the Friedman's rank test, it can be noted that DC consumes the lowest amount of time while RP consumes the highest amount of time for perturbation. But, the values in Table \ref{runtimeanalysis} further suggest that DC and $k'-P^2RoCAl$ lie close to each other in terms of time consumption by the perturbation process.

\subsection{Scalability}

The scalability analysis was conducted using an SGI UV3000 supercomputer (see details in section \ref{expsetup}). The results of scalability experiments with two extremely large datasets (HPDS and HIDS) are given in Table \ref{scalability}. The times to perturb these datasets, having more than 3 million and 11 million records respectively, are shown in the table. As can be seen, RP didn't converge even after 100 hours. 

\begin{table}[H]
	\centering
	\caption{Scalability results (in seconds) of the three methods for high dimensional data }
	\resizebox{1\columnwidth}{!}{
	\begin{tabular}{|l|l|r|r|r|r|}
		\toprule
		\textbf{Dataset} & \textbf{RP} & \multicolumn{1}{l|}{\textbf{DC (k=1000)}} & \multicolumn{1}{l|}{\textbf{DC (k=10000)}} & \multicolumn{1}{l|}{\textbf{$k'-P^2RoCAl$ ($k'$=1000)}} & \multicolumn{1}{l|}{\textbf{$k'-P^2RoCAl$ ($k'$=10000)}} \\
		\midrule
		HPDS  & NC within 100h & 5.34E+03 & 526.1168 & 4.61E+03 & 657.7576 \\
		\midrule
		HIDS  & NC within 100h & 6.95E+04 & 6.42E+03 & 1.36E+05 & 7.28E+03 \\
		\bottomrule
	\end{tabular}%
}
\footnote{Footnote}NC: Did not converge
	\label{scalability}%
\end{table}%

\section{Discussion}
\label{discussion}

In this paper a new data perturbation method ($P^2RoCAl$) was introduced which can be used for both static data and stream data privacy preservation. The proposed method, $P^2RoCAl$ is mainly based on  two data perturbation methods: data condensation (DC) and rotation perturbation (RP). Though DC can be used for both static data and stream data processing, RP can only be used for static data processing. Due to RP preserving the distance between tuples, it often provides good accuracy in data clustering and classification. However, RP is not cost-effective in terms of computer processing. DC, on the other hand, is a fast privacy preservation algorithm which is capable of efficient data stream processing, but its synthetic data generation often decreases the quality of data when it is configured to provide high data privacy. If DC is used under its settings to provide high data privacy (with a larger spatial locality), it often ends up in generating datasets with reduced accuracy. A better method for both static and stream data processing, $P^2RoCAl$ combines the advantages of  DC and RP: high efficiency and high accuracy.

It was noticeable that $P^2RoCAl$ shows increasing accuracy while DC shows decreasing accuracy against an increasing $k'$ value. Especially when the datasets are large (i.e. larger group/cluster sizes and larger spatial locality), we cannot expect good classification accuracy from DC. To have no substantial loss of accuracy, DC assumes that the data is uniformly distributed and has small spacial locality. This leads to reduced privacy, as the distribution of the condensed data is very close to that of original data. As shown in Figure \ref{classificationaccuracy} and Figure \ref{stddifferenceagainstk}, it is noticeable that with increasing $k'$ the accuracy of DC decreases while the $std(D-D^p)$ increases. So, it is clear that for small group sizes, DC provides a higher accuracy while providing lower privacy which is evident from the further analysis on attack resilience, as shown in Table \ref{attackresilience}. In fact, the results available in Table \ref{attackresilience} and Table \ref{friedmanattackranks} further prove that all four derivations of $P^2RoCAl$ provide higher attack resilience than both DC and RP. This proves that DC cannot maintain the balance between privacy and utility on an acceptable level under a single point of convergence. $P^2RoCAl$ shows the opposite effect, where for smaller $k'$ values, it provides higher privacy while providing an accuracy greater than that of DC. This is due to the fact that, as the group size gets smaller, it forms a large number of groups. Therefore, the effect of rotation perturbation with the result of many different angles is greater on the dataset. Since rotation does not distort the distances between tuples, accuracy is not affected. As a result, $P^2RoCAl$ provides better classification accuracy while providing a higher level of deviation of perturbed data from the original data (refer to Figure \ref{classificationaccuracy} and Figure \ref{stddifferenceagainstk}). This feature of $P^2RoCAl$ is essential for data stream perturbation since we have to perform on-demand data perturbation of infinite streams. When perturbing data streams, the resulting number of groups can become unlimited, yet the privacy and accuracy is well preserved by $P^2RoCAl$.

From Table \ref{classyaccuracy} and Figure \ref{averageaccuracy} it is apparent that introducing the k-means algorithm for clustering improves the accuracy of the perturbed dataset. At the same time, k-means decreases the efficiency slightly (see Table \ref{runtimeanalysis}), and resilience suffers (refer Tables \ref{attackresilience} and \ref{friedmanattackranks}). This is due to the fact that k-means algorithm improves the homogeneity of the clusters which results in better accuracy. Since the cluster sizes are different (e.g. some may have even one tuple), perturbation tends to have slightly less effect.

According to Table  \ref{classyaccuracy} and Figure \ref{averageaccuracy}, the derivations of $P^2RoCAl$ provide better accuracy than both DC and RP as reflected in the FMR values shown in the last column of Table \ref{classyaccuracy}. So we can say that $P^2RoCAl$ provides better accuracy for both static and stream cases, than DC and RP. It can be observed that in the static case of data perturbation, $k'-P^2RoCAl\_kmeans$ provides better accuracy compared to that of $k'-P^2RoCAl$. This is because of the increased homogeneity of the groups, which results from  $k-P^2RoCAl\_kmeans$. In $k-P^2RoCAl\_kmeans$, the number of resulting groups is controlled by the parameter $k$. The size of a particular group is not constant;  it uses $k-means$ clustering for grouping. This can result in groups with different sizes, but with higher homogeneity. In $k'-P^2RoCAl$, we control the parameter $k'$, i.e., the group size so that all groups will have a constant number of tuples. Therefore, the groups produced in $k'-P^2RoCAl$, might not be as homogeneous as in $k-P^2RoCAl\_kmeans$. As a result of that, $k-P^2RoCAl\_kmeans$ tends to provide better accuracy. When the method is applied to streams, both $k'-P^2RoCAl\_streams$ and $k-P^2RoCAl\_kmeans\_streams$ seem to provide similar accuracy. This is due to the fact that initial clustering is performed upon the buffered data chunks that are small in size when compared to the main dataset $D$. The small changes in homogeneity will not make a significant impact in changing the accuracy of the final dataset.

Table \ref{classaccuracycomp} and Figure \ref{accuracybox} show the dynamics of classification accuracy for different classification algorithms and different datasets. According to Figure \ref{accuracybox}, $k'-P^2RoCAl$ provides better accuracy than DC in most cases. It can also be noted that the accuracy provided by $k'-P^2RoCAl$ is less than that of RP. This is how $k'-P^2RoCAl$'s multiple rotation perturbations affect the dataset, whereas RP imposes a single perturbation on the whole dataset. As RP is not capable of data stream perturbation, $P^2RoCAl$ can be considered a superior solution.

The time complexity analysis of the proposed algorithm shows that $P^2RoCAl$'s computational complexity is governed by the clustering component. As the number of attributes and tuples increase, the time complexity of the perturbation algorithm will also increase. Conventionally a data stream grows incrementally where new tuples are rapidly added to the dataset while the number of attributes remains constant. Therefore, we can consider an inequality between $m$ and $n$ as $m>>>n$ where $n$ is small compared to an extremely larger number of $m$.  This suggests that the number of tuples has a higher contribution to the time complexity. But, the empirical data on time consumption show that $PR^2CAl$ consumes reasonably low amounts of time and the plots (refer Figure \ref{timecomplexityplots}) on time consumption show patterns close to linear. The very low time consumption of the perturbation process compared to RP was shown in Table \ref{runtimeanalysis}. This further underlines the suitability of $P^2RoCAl$ for fast data streams. Scalability results available in Table \ref{scalability} prove that the proposed algorithm is capable of perturbing big datasets that have extreme dimensions.

\subsection{Selecting values for $k$/$k'$}
We can apply logical reasoning for the selection of suitable values for $k$/$k'$ by looking at the results of $P^2RoCAl$ on different datasets. There is a relative dependency between $l$ and $k/k' $.  The selection of these two input parameters is highly dependent on the speed of a particular data stream.  The faster the data stream (e.g. 4000 tuples per second) , the lower the $l$ value (e.g. 1000). Hence $k'$  needs to be higher (e.g. 200) (or $k$ needs to be lower (e.g. 5)). When the speed of the data stream is relatively slower (60 tuples per second), the vice-versa of the above scenario applies, where $l$ can be relatively higher (e.g. 4000) while $k'$=100 (or $k$=40). But, the selection of specific values for these parameters will not have an extensive effect on the privacy and utility of $P^2RoCAl$ (which is evident in Figures \ref{classaccuracy} and \ref{stddifferenceagainstk}), rather, it will affect the efficiency of the method.

\subsection{Real life application of $P^2RoCAl$: Application Vs. Constraints}
Application of privacy-preserving methods in real-world scenarios involves specific constraints. Most of the applications need attention to particular adaptations. For example, Marcelo Luiz Brocardo et al. discuss how privacy can be imposed over a real world application: a positive credit system. They propose a cryptographic protocol to share private information about customers and companies to produce credit profiles~\cite{brocardo2017privacy}. The proposed version of $P^2RoCAl$ is designed to perturb only numerical data, and real-life applications of $P^2RoCAl$ are limited to numerical big data sets and data streams. Most of the existing IoT sensors (e.g. temperature sensors, proximity sensors, pressure sensors, chemical sensors, IR sensors, etc.) produce numerical data resulting in numerical big data.  This enables $P^2RoCAl$ to be applied to many applications, including, but not limited to, healthcare, financial, social networking, weather, marketing. Some real-world  IoT stream examples that $P^2RoCAl$ is applicable to,  include Sense your City (CITY) and  NYC Taxi cab (TAXI)~\cite{shukla2016benchmarking}. Sense your City (CITY)  is an urban environmental monitoring project that has used crowd-sourcing to deploy sensors at 7 cities across 3 continents in 2015, with about 12 sensors per city, emits  7000 messages/ sec.  NYC Taxi cab (TAXI)  offers a stream of smart transportation messages that arrive from 2M trips taken in 2013 on 20, 355 New York city taxis equipped with GPS emits 4000 messages/ sec.  Empirical analysis of $P^2RoCAl$ shows that the perturbation explicitly preserves the classification properties of the original data. Therefore, the proposed method is best utilized for privacy-preserving data classification.

\subsection{Future directions}
The proposed algorithm can be configured via different permutations of settings. Part of the algorithm configuration was explained in Section \ref{configurations}. $P^2RoCAl$ configuration can also be changed with the two other factors: the distance measure for homogeneous clustering and the method used for the matrix decomposition (Equation \ref{eiequ}). We have the possibility of using several distance measurements for the random data grouping method as well as for $k-means$ clustering. As the proposed algorithm can be used for any numerical dataset, we should avoid using binary distance measurements such as "hamming distance" \cite{banka2015hamming}. The results provided in this paper are based on $P^2RoCAl$ that uses $Euclidean$ distance \cite{cha2007comprehensive} as the distance measure. Some other distance measurements that can be effectively used with $P^2RoCAl$ are City block distance \cite{cha2007comprehensive}, Cosine distance \cite{cha2007comprehensive} and Correlation \cite{cha2007comprehensive}. $P^2RoCAl$ uses eigenvalue decomposition of the covariance matrix generated by each group. Other decomposition methods, such as singular value decomposition, polar decomposition, etc. can also be used to generate the rotation matrix of a particular group, as long as the decomposition generates a sub-matrix that has the properties of a rotational matrix. Further studies on increasing the efficiency of $P^2RoCAl$ using sampling techniques and parallel implementations can be investigated. This would allow $P^2RoCAl$ to work with high-speed data streams.
 
\section{Conclusion}
\label{conclusion}
A new data stream perturbation algorithm ($P^2RoCAl$) was introduced. It provides higher accuracy, efficiency and attack resilience than similar methods. It was shown that the runtime complexity is governed by clustering when the number of attributes is kept constant. The algorithm shows a worst-case runtime complexity of $O(n^3)$ when the number of tuples is maintained as a constant.  As the number of attributes is very low compared to the number of tuples in many cases, e.g. in data streams, $P^2RoCAl$ exhibited a considerably lower time consumption during the empirical analyses. This makes it possible to work with continuously growing data streams and big data. The proposed method  $P^2RoCAl$ showed better classification accuracies than its contenders.  $P^2RoCAl$ also shows higher resilience against the attacks such as naive estimation, I/O attacks, and ICA attacks, compared to rotation perturbation and data condensation.

In summary, this paper proposed $P^2RoCAl$, an effective perturbation method for data streams and big data. One potential application of $P^2RoCAl$ might be the precision health domain where a large number of IoT devices are or will be used to monitor a person’s body, activities, and behaviors. $P^2RoCAl$ can be used effectively to perturb continuous streams of data generated by sensors monitoring an individual or group of individuals before the data is transmitted to cloud systems for further analysis.

\end{document}